\documentclass[letterpaper,twocolumn,10pt]{article}
\usepackage{usenix-2020-09}
\usepackage{tikz}
\usepackage{amsmath}

\usepackage{filecontents}
\usepackage{multirow}
\usepackage[most]{tcolorbox}
\usepackage{float}

\usepackage{xcolor}
\usepackage{multirow}
\newif\ifcomment
\commentfalse
\ifcomment
\newcommand{\shirin}[1]{{\bf \textcolor{purple}{Shirin: #1}}}
\newcommand{\sayak}[1]{{\bf \textcolor{blue}{Sayak: #1}}}

\newcommand{\elham}[1]{{\bf \textcolor{teal}{elham: #1}}}
\else
\newcommand{\shirin}[1]{}
\newcommand{\sayak}[1]{}
\newcommand{\elham}[1]{}
\fi

\begin{document}
%
% paper title
% can use linebreaks \\ within to get better formatting as desired
\title{DarkGram: A Large-Scale Analysis of Cybercriminal Activity Channels on Telegram }

%for single author (just remove % characters)
\author{
  Sayak Saha Roy \\
  Louisiana State University  \and
  Elham Pourabbas Vafa \\
  University of Texas at Arlington   \and
  Kobra Khanmohammadi \\
  Sheridan College 
  \and
  Shirin Nilizadeh \\
  University of Texas at Arlington \\
}
\maketitle

\begin{abstract}
%\boldmath

We present the first large-scale analysis of 339 cybercriminal activity channels (CACs). 
Followed by over 23.8M users, these broadcast-style channels share a wide array of malicious and unethical content with their subscribers,  including compromised credentials, pirated software and media, social media manipulation tools, and blackhat hacking resources such as malware and exploit kits, and social engineering scams. 
To evaluate these channels, we developed DarkGram—a BERT-based framework that automatically identifies malicious posts from the CACs with an accuracy of 96\%. Using DarkGram, we conducted a quantitative analysis of 53,605 posts posted on these channels between February and May 2024, revealing key characteristics of shared content. 
While much of this content is distributed for free, channel administrators frequently employ strategies, such as promotions and giveaways, to engage users and boost the sales of premium cybercriminal content. 
Interestingly, sometimes, these channels pose significant risks to their own subscribers. 
Notably, 28.1\% of the links shared in these channels contained phishing attacks, and 38\% of executable files were bundled with malware. 
Looking closely into how subscribers consume and react positively to the shared content paints a dangerous picture of the perpetuation of cybercriminal content at scale. 
We also found that the CACs can evade scrutiny or platform takedowns by quickly migrating to new channels with minimal subscriber loss, highlighting the resilience of this ecosystem. 
To counteract this, we utilized DarkGram to detect emerging channels and reported malicious content to Telegram and the affected organizations. This resulted in the takedown of 196 channels over the course of three months. 
Our findings underscore the urgent need for coordinated efforts to combat the growing threats posed by these channels. To aid this effort, we open-source our dataset and the DarkGram framework.

\end{abstract}

\section{Introduction}

Underground online forums have long served as hubs for cybercriminal activity, where attackers exchange resources for executing social engineering scams~\cite{hao2015drops,herley2010nobody}, coordinating malware attacks, sharing leaked user credentials and distributing hacking tools~\cite{pastrana2019first,marin2018community}. 
These forums are typically dominated by a small group of experienced members who share a range of malicious resources~\cite{afroz2013honor}. 
However, combined efforts by law enforcement, security vendors, and researchers towards monitoring, infiltrating, and shutting down these platforms—especially those on the clear web—have reduced the impunity with which cybercriminals operate~\cite{bada2021understanding}. 
Consequently, many cybercriminals have turned to alternative platforms such as social media, which provide a more dynamic environment for exploiting vast user bases while evading traditional security measures~\cite{elezaj2021criminal}.

One platform that has emerged as a new frontier for cybercrime is Telegram~\cite{Telegram}, a messaging service. % known for its encrypted communications.
Recently, Telegram has faced criticism for its lax content moderation policies~\cite{BBCTelegram, TheNewYorkerTelegram}. With over 700 million monthly active users, it has become an attractive platform for cybercriminals seeking a broad, accessible audience\cite{la2021uncovering}. 
Unlike dark web forums, which require specialized software\cite{biswas2022text}, Telegram channels can be publicly accessible, lowering the barrier of entry for aspiring cybercriminals and those looking for tools and resources to engage in malicious activities.
These Telegram channels mirror the behavior seen in underground forums, enabling large-scale distribution of illegal content. 
Our paper offers the first large-scale study of this cybercrime ecosystem, examining 339 dedicated \emph{Cybercriminal Activity Channels (CACs)} on Telegram.
These channels are used to distribute compromised account credentials, pirated media, and software, social media manipulation tools, and hacking resources—including exploits, social engineering kits, and malware. 
We characterize both the content shared within these channels and how subscribers engage with it. 

Our paper is structured as follows:
Section~\ref{data_collection_and_processing} outlines our methodology for identifying channels within five distinct CAC categories, as well as our processes for collecting and pre-processing data (posts and their metadata) from these channels. 
Section~\ref{research_methodology} describes the five stages of our analysis: 1) a manual review of a random sample of posts to characterize content shared in each CAC.  %(Section~\ref{characterization})
2) Then using these manually labeled samples to train a multi-class BERT model, DarkGram, which classifies posts based on their text with an average accuracy of 96\% across different CAC categories, allowing us to extend our findings across the entire dataset. 
This leads to 3) Section~\ref{characterization}, where we conduct an in-depth analysis of each CAC category, examining the characteristics and volume of shared content, the methods of distribution, key stakeholders affected by this content, and our efforts to alert them of these threats to enable appropriate action. 
4) In Section~\ref{harm_analysis}, we focus on the potential harm caused by CAC posts, not only to organizations and individuals targeted by the content but also to the channel subscribers themselves.
We do so by looking for phishing scams, malware, and other fraudulent activities within the shared content. 
This is particularly important given the public nature of these Telegram channels, which makes harmful content accessible to novice users who may unknowingly download malicious content.
5) Then, in Section~\ref{engagement_analysis}, we investigate user engagement towards the shared content through metrics, such as post views, forwards, and emoji reactions, to understand how audiences interact with the content and assess their satisfaction with the content. 
In Section~\ref{sec:interpretation}, we contextualize our findings by examining the cross-platform dynamics between Telegram CACs and traditional cybercriminal forums. We highlight key behavioral and structural differences while uncovering novel insights that underscore the unique role of Telegram CACs.  
% within the broader cybercriminal ecosystem, 
Finally, in Section~\ref{darkgram}, we run our DarkGram in real-time on Telegram and Facebook to identify new CACs and disclose them to Telegram and various stakeholders— an effort in the takedown of 196 such channels over nearly three months. 
% To further support the research community and ongoing security efforts, we open-sourced both our dataset and the DarkGram model at \url{https://tinyurl.com/mjt38z53}. 
The primary contributions of our work are as follows:

\begin{enumerate}
    \item We monitored 339 Telegram channels, collectively having over 23.8M subscribers, dedicated to sharing cybercriminal content across five categories: 1) Compromised user credentials, 2) Pirated software, 3) Pirated media, 4) Social media manipulation tools, and 5) Blackhat hacking resources. Our analysis %, conducted between February 21 and May 29, 2024, 
    uncovers key characteristics of the shared content and highlights various distribution strategies, including payloads directly uploaded to the platform, the use of bots, and external links.

    \item We developed \textit{DarkGram}, a BERT-based framework that automatically detects malicious posts with an average accuracy of 96\% across the five categories. This enabled us to identify and analyze 53,605 posts from the 339 CACs and further helped us takedown 196 new CACs shared on Telegram and Facebook.

    \item Through an analysis of user engagement with the posts, we reveal not only the high activity levels of these channels but also the diverse strategies channel administrators employ to foster trust among subscribers. Our findings indicate that subscribers respond positively to the content, with nearly 78\% of emoji reactions expressing approval. Moreover, CACs exhibit notable resilience to takedown efforts, rapidly relocating subscribers to new channels. % in response to potential threats.

    \item We found that CACs also post content which \emph{can be malicious to the subscribers}, with 28.1\% of the posts leading to phishing URLs and 38\% of shared executables containing malware. 
    Despite this, the majority of emoji reactions to these posts remained positive, suggesting subscribers are oblivious to the content's real nature. 

    \item We contextualized our findings within the existing literature on traditional cybercriminal forums, highlighting how Telegram has emerged as a powerful platform for distributing malicious and illegal content.

    \item Finally, to further aid the research community in identifying malicious CACs, we have open-sourced our dataset, the first of its kind, along with the DarkGram framework, available at \href{https://github.com/Scalable-Security-Research-Lab/DarkGram}{https://github.com/Scalable-Security-Research-Lab/DarkGram}.

\end{enumerate}

\section{Data Collection and Pre-processing}
\label{data_collection_and_processing}
\subsection{Identifying seed channels}
\label{identifying-seed-channels}
To compile a list of channels sharing cybercriminal content, we utilized Telemetr.io~\cite{telemetr2024}, a third-party catalog of active, public Telegram channels. Unlike the official Telegram API—which has limited global search functionality, Telemetr.io indexes relevant channels across several categories using keyword-based crawling, user submissions, and automated tracking of publicly available metadata. For each channel, Telemetr.io displays a range of metrics, such as subscriber counts, engagement rates, and historical posting data. This functionality allowed us to discover 4,709 English-based channels with 10,000 or more followers.   
We focused on channels with a higher number of subscribers as they are likely to be well-established, with several posts, allowing us to gain a comprehensive understanding of the breadth of cybercriminal activity on these channels. 
We acknowledge that this approach might have introduced potential biases by omitting smaller or newly emerging channels. Despite this limitation, our seed dataset is crucial for training DarkGram (Section~\ref{darkgram}), which was able to identify newer and smaller CACs. %and subsequently compare these newly detected channels to their more established counterparts.

To identify the channels' purposes, two coders manually reviewed each channel's description text and ten of its latest posts.
Our criteria for identifying malicious activities were derived from the Federal Bureau of Investigation's Internet Crime Report (2023)~\cite{ic32023report}, which identifies cybercrime categories, such as \emph{identity theft}, \emph{personal data breach}, \emph{copyright infringement}, \emph{malware}, \emph{phishing}, \emph{system exploits}, etc. Overall, we identified 339 channels that exhibited one or more of these characteristics \textit{at least once} in their first 10 posts. 

Based on the characteristics of these channels, we streamlined the categories that we study into the following \emph{five}:
\textbf{1)~Credential compromise} channels provide leaked/ hacked account credentials obtained from various online services.
\textbf{2)~Pirated software} channels distribute software without authorization. This includes modded versions of paid software allowing the users to bypass legal purchase methods, \textbf{3) Blackhat resource} channels distribute tools and resources used for committing cybercrime, including malware, hacking tools, scripts, and other software designed to facilitate illegal activities online, \textbf{4) Pirated Media} channels distribute unauthorized or unlicensed media content, such as movies and TV shows, and finally, \textbf{5) Social media manipulation} channels provide services to inflate social media metrics artificially, including selling likes, followers, and engagement.
% to individuals and businesses looking to boost their online presence and credibility illicitly.

\subsection{Collecting posts} To gather posts from these 339 channels, we used the official Telegram API~\cite{telegramAPI} to collect 64,801 posts between February 21st and May 29th, 2024. Each channel was queried at 10-minute intervals to capture new posts while adhering to the API rate limits. 
We also monitored metrics such as subscriber count, post views, and forwards at 10-minute intervals. Telegram channels, by design, only allow channel admins to post original content and decide whether to enable user replies~\cite{telegram_channels}. Therefore, we collected replies to posts whenever they were present. Among 339 channels, 167 allowed replies to their posts. Lastly, we collected the emoji reaction to posts, as they serve as a good indicator of how the user resonates with the content~\cite{ling2021dissecting}. 
To ensure our analysis was based on the most current data, we refreshed (recollected) each post at 10-minute intervals, capturing the latest counts for views, forwards, emoji reactions, and reply content. % throughout the study period. 

\begin{table*}[]
\centering
\resizebox{0.9\textwidth}{!}{%
\begin{tabular}{c|cc|cc|cc|cc}
\hline
\multirow{2}{*}{Channel category} & \multicolumn{2}{c|}{Posts} & \multicolumn{2}{c|}{Subscribers} & \multicolumn{2}{c|}{Views} & \multicolumn{2}{c}{Forwards} \\ \cline{2-9} 
 & \multicolumn{1}{c|}{Min/Max} & Median/Mean & \multicolumn{1}{c|}{Min/Max} & Median/Mean & \multicolumn{1}{c|}{Min/Max} & Median/Mean & \multicolumn{1}{c|}{Min/Max} & Median/Mean \\ \hline

Credential Compromise (69) & \multicolumn{1}{c|}{1/3159} & 102/520 & \multicolumn{1}{c|}{10,215/604,772} & 28,500/47,132.25 & \multicolumn{1}{c|}{1/19,300} & 23/65.99 & \multicolumn{1}{c|}{0/483} & 0/0.39 \\ %\hline
Pirated Software (124) & \multicolumn{1}{c|}{1/841} & 63.5/95.16 & \multicolumn{1}{c|}{11,030/741,259} & 31,200/52,415.89 & \multicolumn{1}{c|}{1/93,627} & 397/3138.78 & \multicolumn{1}{c|}{0/504} & 1/6.58 \\ %\hline
Blackhat Resources (42) & \multicolumn{1}{c|}{3/290} & 35.5/60.5 & \multicolumn{1}{c|}{10,481/253,507} & 18,450/29,763.11 & \multicolumn{1}{c|}{1/44,573} & 259/1079.25 & \multicolumn{1}{c|}{0/597} & 1/7.75 \\ %\hline
Pirated Media (36) & \multicolumn{1}{c|}{2/400} & 25/56.64 & \multicolumn{1}{c|}{15,902/387,599} & 60,000/102,347.89 & \multicolumn{1}{c|}{1/796,961} & 24,395.5/47,027.21 & \multicolumn{1}{c|}{0/2569} & 14/89.03 \\ %\hline
Social media manipulation (68) & \multicolumn{1}{c|}{1/1227} & 8/49.15 & \multicolumn{1}{c|}{10,319/123,097} & 20,750/28,321.44 & \multicolumn{1}{c|}{1/40,757} & 326.5/753.78 & \multicolumn{1}{c|}{0/117} & 0/5.01 \\ \hline
\end{tabular}}
\caption{Descriptive statistics of the Cybercriminal Activity Channels (CAC)}
\label{descriptive-stats}
\end{table*}

%%%%%%%%%%%%%%%%%%% SECTION 4 %%%%%%%%%%%%%%%%%%%%%%%%%

\section{Research Methodology}
\label{research_methodology}

\begin{figure*}[t]
\centering
  \includegraphics[width=0.75\textwidth]{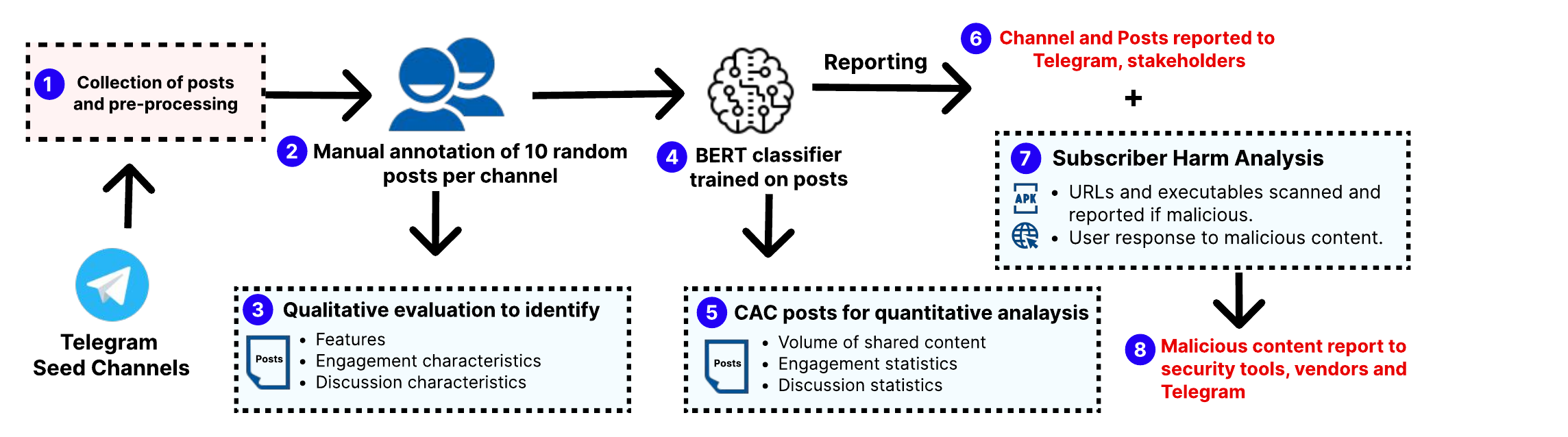}
\caption{DarkGram framework for identifying and reporting new cybercriminal activity channels}
  \label{darkgram-framework}
\end{figure*}

To comprehensively study CACs, our paper is structured around the key phases illustrated in Figure~\ref{darkgram-framework}: 

\textbf{Characterizing the CACs:} Our dataset consists of 64,801 unique posts across 339 channels, making it impractical to qualitatively analyze such a large number of posts. 
Therefore, we randomly selected 10 posts from each channel, resulting in a total of 3,390 posts, which were manually evaluated by two independent coders who were graduate students in Computer Science. 
Coder 1 specialized in Computer Security and Social Computing, whereas Coder 2 had good experience in Computer Security through research and academic coursework.  
Nearly 91.4\% of posts (n=3,098) were found to be distributing cybercriminal content.
The coders' Cohen's Kappa interrater agreement was 0.78, suggesting substantial agreement and the disagreements were resolved. 
Note that the posts evaluated for this process are exclusive to the ones we had already used to identify the channels from Telemetro to build our seed dataset in Section~\ref{identifying-seed-channels}. 
We believe that sampling posts in this manner provided the best representation of the content shared in these channels, preventing bias towards more recent posts and offering a more balanced and comprehensive view of the content. 
In Section~\ref{characterization}, we qualitatively examined these posts to identify the characteristic features of content shared in the channels across all five CAC categories.
% , highlighting different strategies used to distribute cybercriminal content and the characteristics of the content itself.  
Building on these insights, we further extended our evaluation quantitatively to a larger sample of such posts, automatically identified by our classifier. 

\textbf{Automating detection and Quantitative Analysis:} 
To perform a quantitative analysis across our entire dataset and, later on, to automate the detection and reporting of new CACs to aid in their removal, we trained a classifier on the text content of the posts.
We used the 3,098 posts manually labeled as malicious in our groundtruth for \emph{True} labels. 
For \emph{False} labels, we sampled the same number of posts from the Pushshift Telegram~\cite{baumgartner2020pushshift}, a dataset of over 317M messages collected from 27.8K Telegram channels. 
We manually verified these posts to ensure they contained no cybercriminal content.

Our model first focuses on identifying whether a post is a Cybercriminal Activity (CA) post and, if so, determining which CAC category the post belongs to.
We split these labeled samples into a 70:30 training-test split and trained a BERT-base model on the post text, which returned an F1-score of 98.4\%. 
We chose this model since language models like BERT have demonstrated excellent performance on text classification tasks~\cite{kaliyar2021fakebert,acheampong2021transformer}, as they are adept at capturing nuanced patterns and contextual relationships within text, such as those found in the Telegram CAC posts. 
We opted not to use commercial LLMs such as ChatGPT due to concerns about data privacy, API costs, and speed.
% for both fine-tuning and inference.
Running the model on our full dataset, it identified 53,605 posts (About 83\% of the full dataset) as sharing content relevant to a respective CAC category. 
To further evaluate the model's efficiency, our coders labeled an additional 1,000 randomly selected posts marked as CAC positive (200 posts per CAC category, mutually exclusive from the previously sampled 3,098 posts used for training the model) and 1,000 posts marked benign by the model. 
The model achieved an F1 score of 97.8\%. % across these samples. 
Table~\ref{table:performance_metrics} illustrates the overall performance of the model over the randomly selected posts, as well as performance per CAC category, suggesting that it performs exceptionally well across all categories, except for the Pirated Software category. 
% It is important to note that nearly 32\%  of the posts in this category were files shared without any accompanying text. 
Likely, the absence of textual content (since 32\%  of the posts in this category are files without text)  contributed to the lower performance, as the model relies heavily on textual information.
\newline
\begin{table}[]
\centering
\resizebox{\columnwidth}{!}{
\begin{tabular}{l|c|c|c|c}
\hline
Category (200 per) & Accuracy & Precision & Recall & F1-score \\ \hline
Credential Compromise & 98.1\% & 98.3\% & 97.9\% & 98.1\% \\ 
Pirated Software & 88.7\% & 89.2\% & 88.3\% & 88.7\% \\ 
Blackhat Resources & 97.4\% & 97.2\% & 97.6\% & 97.4\% \\ 
Pirated Media & 99.2\% & 99.0\% & 99.4\% & 99.2\% \\ 
Social-media Manipulation & 96.2\% & 95.8\% & 96.6\% & 96.2\% \\ 
Overall & 97.8\% & 97.9\% & 97.7\% & 97.8\% \\ \hline
\end{tabular}}
\caption{Performance of BERT-based detection model for different CAC categories}
\label{table:performance_metrics}
\end{table}
\textbf{Harm to subscribers:} Traditional cybercriminal forums, while serving as hubs for sharing malicious content, are also rife with bad actors who exploit their peers by sharing scams or harmful material that directly targets forum members~\cite{yip2017trust,holt2016examining}. 
Section~\ref{harm_analysis} investigates whether CAC administrators 
% (who have significantly more power and influence over the visibility of the content than typical forum members) 
exploit the trust of their audience by sharing content that can be harmful to the subscribers. We systematically use tools, such as VirusTotal and Hybrid Analysis, to identify phishing scams and malware disguised within the shared content.

\textbf{Engagement Analysis:} Building upon the characterization of CACs in Section~\ref{characterization}, we dedicate Section \ref{engagement_analysis} towards identifying the intricate dynamics of engagement within these cybercriminal communities on Telegram. 
We examine how channel administrators utilize strategies to attract and retain subscribers, including sharing proofs, giveaways, or frequent updates, and how these tactics influence subscriber growth over time. 
Simultaneously, we investigate how subscribers interact with the posted content, evaluating explicit engagement metrics, such as post forwards and subscriber growth, and implicit feedback through emoji reactions. 
%%%%%%%%%%%%%%%%%%%%% SECTION 5 %%%%%%%%%%%%%%%%%%%%%%%%%%%%%
\section{Characterization of CACs}
\label{characterization}
This section presents a detailed analysis of the content and engagement patterns in channels across each CAC category. 
This analysis is informed by a qualitative evaluation of 10 randomly selected posts from each channel and the subsequent extrapolation of these findings to a much larger dataset of posts identified using our automated classifier. %, as described in Section~\ref{research_methodology}.
% (``Automating Detection and Quantitative Analysis''). 

\subsection{Credential compromise channels}
\label{credential_compromise}
\begin{figure}[t]
\centering
  \includegraphics[width=0.6\columnwidth]{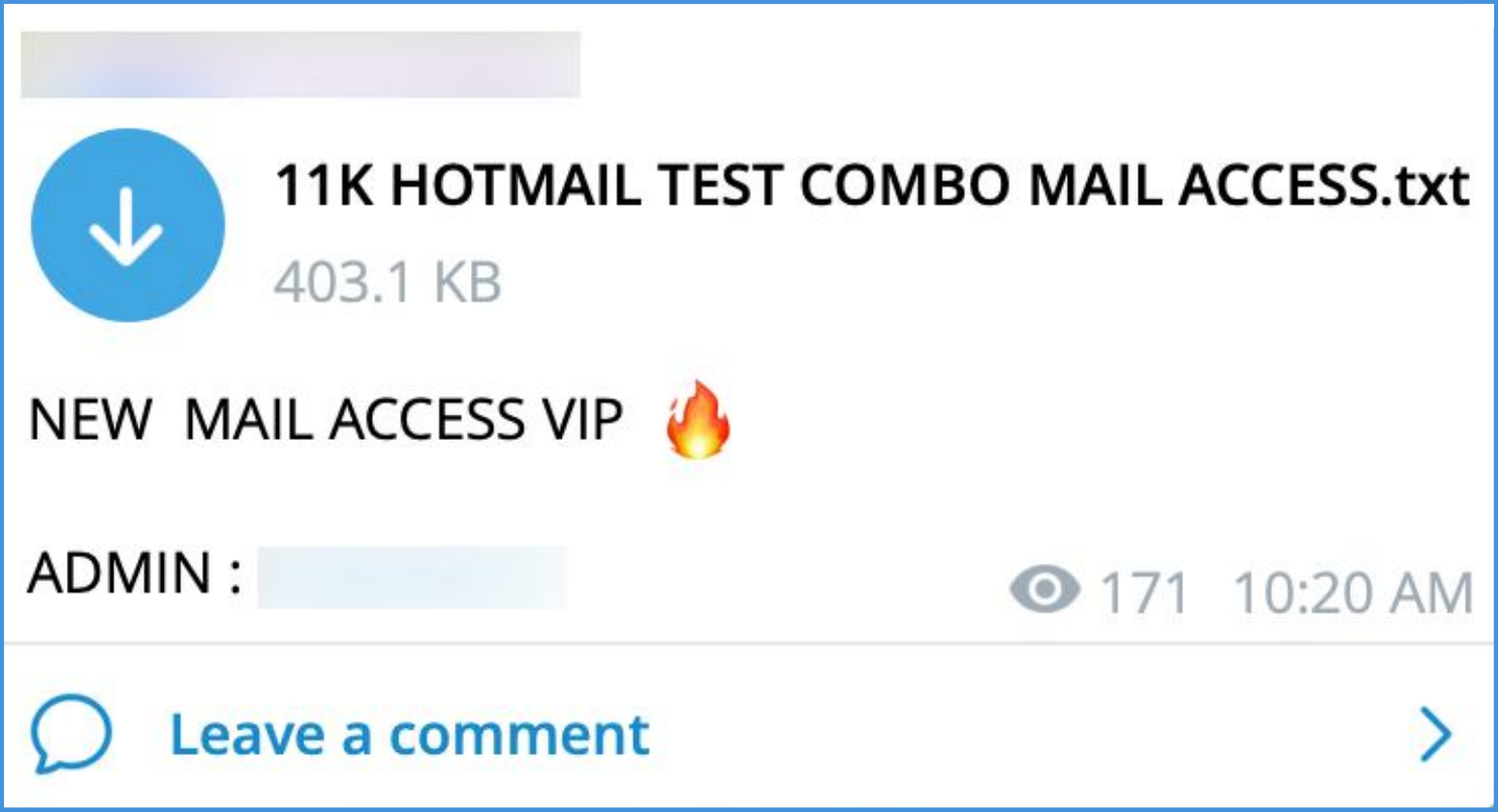}
 \caption{Example of a payload file shared in a Credential Compromise channel}
  \label{fig:credential_blackhat}
\end{figure}

We manually evaluated 690 posts across 69 channels that claimed to share account-compromised user credentials and session cookies for various online services.
We found that posts in these channels typically followed a consistent structure. First, they mentioned the entity or organization whose credentials had been leaked, followed by a file attachment that acted as the payload, i.e., containing the leaked credentials/session cookies or an external link leading to a website with the payload. 
Figure~\ref{fig:credential_blackhat} illustrates one such post that shares 11,000 Hotmail account credentials in a text file.
Our automated classifier further identified 35,883 posts across these channels.
For ethical reasons, due to the presence of personally identifiable information, we did not download or read the payload files. 
Instead, our proceeding statistical analysis is solely based on the information provided in the post text and the payload file names. 
It was also essential to distinguish between posts sharing user credentials and those sharing session cookies, as these may require different handling and mitigation strategies from a security vendor's point of view~\cite{wagner2020cookies}. 
% Since our automated classifier cannot differentiate between these two types of attacks, 
We utilized spaCy's~\cite{spacy_github} tokenization to identify key phrases and patterns in the post content and filenames. 
This approach was chosen over simple word frequency as spaCy systematically breaks down text into meaningful components while capturing contextual patterns and semantic relationships. 
This was particularly valuable in cases where posts lacked descriptive text and relied solely on filenames of attached files to infer their content (25.8\% of posts). 
Using this information, we developed a regular expression that was capable of accurately distinguishing between posts sharing user credentials and those sharing session cookies.
To check the efficiency of this approach, we manually checked 100 random samples that were predicted as sharing user credentials and 100 samples that were predicted as session cookies, finding it to be accurate for all cases in the former and all but one case in the latter.
Overall, we found 87.9\% of the posts sharing user credentials, whereas (12.1\%) shared session cookies. 
\newline
% \newline
\textbf{Payload file attachments:} Approximately, 54.2\% (n=19,457) of the posts had file attachmement payloads. 
According to the file names, about 18.7\% of these files contained 10,000 or more credentials, and in 27.4\% of cases, credentials for the same service were exploited across multiple countries (e.g., Gmail accounts in the US, UK, and other countries). 
\newline
\textbf{External links:} 14,574 posts included external links that directed users to websites hosting the payloads. While the former could be analyzed using the post metadata itself, for the latter, we had to deploy a Selenium-based crawler that interacted with the websites to get the webpage title (that can indicate the target of the leak) and to click on links on the website to invoke a file download and extract the file name (without actually saving the file). We then ran our regular expression to identify credential and session cookie files from files downloaded from 89\% (n=12,998) of these links, with the rest being found (upon manual inspection) to either being inactive/dead or having links hidden behind paywalls, necessitating upfront payments to access the files. These paywalls employed various monetization strategies, such as cryptocurrency payments or prepaid card codes, making further automated analysis infeasible.
\newline
\textbf{Bot interactions:} We found 1,852 posts across 64 channels that had tagged other Telegram accounts. Upon manual inspection, we observed that these posts either shared a small sample of leaked credentials or just the description of the leak and encouraged users to contact a bot or user to access or purchase the complete leak. For ethical reasons, we did not engage directly with the users; however, we analyzed the characteristics and behavior of the bots associated with these posts. We include this analysis in in the Appendix~\ref{bot_interactions}.
\textbf{Proofs and guidance:} We also found that 29.2\% of the posts contained the keyword ``proof/proofs'', aimed at building trust and credibility among potential buyers. Manually evaluating a random sample of 1,000 such posts, we found that 849 of them contained screenshots or videos showing successful logins or transactions, serving as evidence of the authenticity of the leaked credentials. Out of them, 173 posts also included instructions on using the credentials, which were sometimes more complex than simply entering them into login fields. For example, some provided step-by-step guides on importing session cookies into a browser to bypass login procedures. In 9.4\% of cases, the posts also contained warnings about potential risks and guidance on avoiding detection by security systems, such as advising against using the same credentials across multiple IP addresses or accessing the accounts from known devices or networks to avoid linking back to the user. In a small number of cases (5.2\%), channels asked subscribers to verify if the shared credentials worked, leading to discussions about their effectiveness and subsequent posts of verified, working credentials. This crowd-sourced validation not only distributed the risk of detection but also reinforced trust in the community, as verified credentials could be used by other members with greater confidence.
\newline
\textbf{Disclosure:} We reported the posts and the respective CACs to Telegram and the targeted organizations. This included 32 email providers, 11 domain providers, and 24 online services, all through their respective vulnerability disclosure programs, starting from the first week of April when we manually reviewed our data and identified the threats. Verifying compromised account credentials can be time-consuming and may require manual evaluation by the respective Incident Response Teams. To date, 15 email providers, six domain providers, and 11 online services have confirmed that many credentials from our report had indeed been breached. Additionally, four email providers and one online service classified our report as ``High Priority,'' while two domain providers promised affirmative action based on our report.

\subsection{Pirated software}
\label{pirated_software}

\begin{figure}[t]
\centering
  \includegraphics[width=0.8\columnwidth]{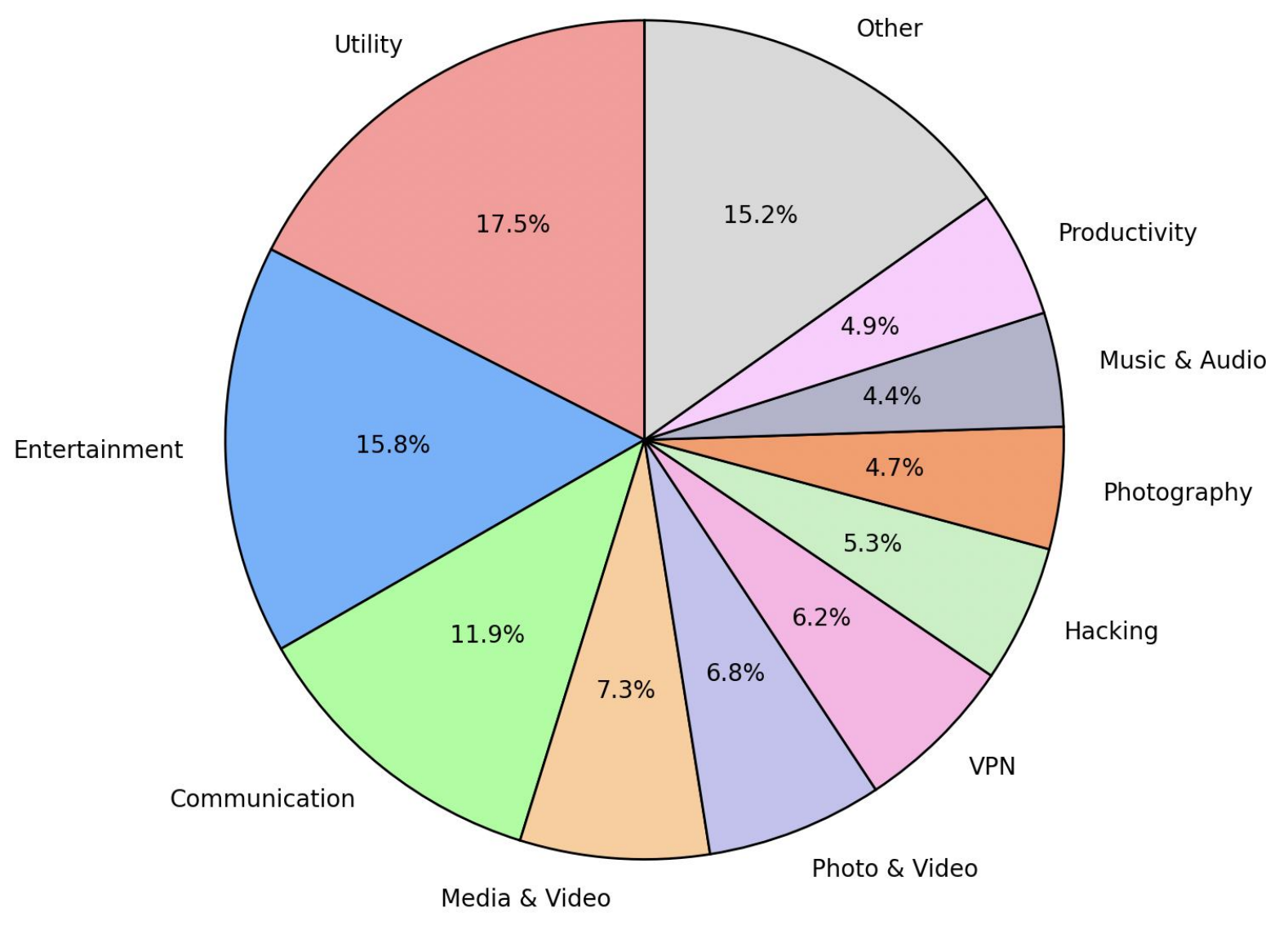}
\caption{Top 10 categories of applications distributed in the Pirated Software CACs}
  \label{fig:pirated_software}
\end{figure}

We manually evaluated 1,240 posts across 124 channels that were dedicated to sharing pirated software and found them sharing modified versions of popular applications for Android and Microsoft Windows across 15 distinct categories. 
Distributing pirated software/media is a serious offense that is actively banned by law enforcement through intellectual property rights~\cite{Greco2023}. Engaging in or facilitating the distribution of pirated content can lead to severe legal consequences, including fines and imprisonment~\cite{hill2007digital,LETRAN}. 
Our classifier further identified 11,800 such posts across these channels. 
% % Unlike credential compromise CACs, 
% We did not find any instances where the channels asked for payment to give access to the software, likely reflecting the primary intention of users visiting these channels—to obtain free software. 
\newline
To extend the software categorization across all the 11,800  posts identified by the classifier, we utilized the GPT-4 API~\cite{openai_azure} to automatically classify files based on the post text and the filename (when an attachment was present) within the existing 15 categories, with the potential to identify additional categories. 
We include the prompt used for this categorization in Section~\ref{prompt_categorization} in the Appendix.
Using this approach, GPT-4 added four new categories, bringing the total number of categories to 19. Our coders validated the post category and description for a random sample of 1,000 such posts and found 994 to be accurate. Figure~\ref{fig:pirated_software} illustrates the ten most common categories of software that appeared in these channels. Using GPT-4, we also identified that the 11,800 posts shared 6,154 applications, with 4,201 being for Android and 1,953 for Windows.
In the majority of the posts (93.8\%, 11,074), the software was shared directly as a file upload, though 2,144 of them also contained additional website links. Evaluating 100 of these samples manually, we found that these extra links were  "backups" on an external website or cloud service. The remaining 726 posts did not have a file upload but relied on an external link (417), which also contained backup links. For example, a message might include multiple links, such as "Download link: [Shortened URL]. Backup link: [Another URL]." This redundancy ensures that users can access the software even if one link is taken down. Finally, for 309 posts, the users were directed to interact with one of 51 unique bots that had characteristics similar to those found previously in Credential Compromise CACs, and discussed in  Section~\ref{bot_interactions} in the Appendix.
\newline A notable percentage (32.4\%) of the posts did not provide any file descriptions, instead only adding the file attachment (similar to some posts in the Credential compromise CACs). We assume that this lack of detail can be attributed to the primary objective of these channels of facilitating quick and easy access to unauthorized software. Channel administrators might assume that their audience is already familiar with the software such that the name of the file already provides sufficient information. Moreover, minimizing descriptions may be a tactic to evade detection and reduce the risk of content being flagged or removed by Telegram, as content moderation systems struggle to detect posts with sparse text content~\cite{huertas2023countering,wang2023mttm}. 
Conversely, using the post description from GPT-4, we found that ~19\% of them included installation and usage instructions which can be particularly valuable for users unfamiliar with the software. Additionally, these detailed descriptions can help build trust and credibility with the audience, particularly for users hesitant about downloading software from an unofficial source. 
\subsubsection{Financial Damage through Pirated Software CACs}
Distributing pirated software not only infringes intellectual property~\cite{goel2009determinants,danaher2014gone} but also financially harms the legitimate developers and organizations who create and maintain these applications. To quantify the financial damage caused by these CACs, we again utilized GPT-4 to estimate the expected price of such software. Since a significant portion of the software was designed for Android, we also tasked GPT-4 with categorizing the software into two pricing models which are dominant in Android app stores~\cite{nieborg2016premium,deubener2016typology}:  Freemium apps that require a subscription/purchase to access features or remove ad-supported and Premium apps that require an outright purchase to access and use. These categories are applicable to popularly pirated Windows software as well~\cite{hsu2008consumers}. We evaluated 100 randomly selected cases and found GPT-4's predictions to be accurate for 97 of them. 
In total, we identified 2,642 freemium-based apps and 3,512 premium apps, with prices ranging from less than a dollar to as high as \$5,000. The median price of these premium apps was \$4.99. We evaluated 100 apps from each of the two categories on a Google Pixel 3 smartphone (for Android apps) and a Windows 11 Virtual Machine (VMWare, for Windows apps), and our findings revealed that the appeal of pirated software varied based on its pricing model. For freemium apps, users were primarily drawn to modifications that removed ads, unlocked additional levels, or provided in-game currency. For subscription-based apps, piracy was largely motivated by the ability to bypass paywalls and access subscription content for free. In contrast, the primary driver for pirating premium apps was straightforward: obtaining paid software without cost.

We estimated the financial losses incurred by developers based on the number of users downloading and using pirated software from these CACs. Since Telegram does not provide download statistics, we used view counts as a proxy to estimate software usage. Table~\ref{cost_analysis} presents our analysis, assuming a conservative estimate where only 10\% of the views translate into actual downloads/usage. Even under these modest assumptions, our findings indicate that developers \textbf{potentially lost over \$40 million} due to apps shared on these CACs.

\begin{table}[]

\centering
\resizebox{0.95\columnwidth}{!}{
\begin{tabular}{c|c|c|c|c|c}
\hline
\textbf{Category (Count)} & \textbf{Min} & \textbf{Max} & \textbf{Median} & \textbf{Mean} & \textbf{10\% conversion} \\ \hline \hline
Communication (732) & \$0.1 & \$50 & \$4.99 & \$6.45 & \$1,115,436 \\ 
Education (8) & \$0.99 & \$399.99 & \$9.99 & \$32.79 & \$1,363,305 \\ 
Entertainment (972) & \$0 & \$2,099 & \$5.99 & \$28.91 & \$4,334,078 \\ 
Finance (31) & \$0.2 & \$623 & \$4.99 & \$35.00 & \$178,439 \\ 
Food \& Drink (141) & \$2.99 & \$3.99 & \$3.49 & \$3.49 & \$1,974 \\ 
Gaming (82) & \$0 & \$99.99 & \$4.99 & \$8.22 & \$1,362,768 \\ 
Hacking (326) & \$0 & \$999 & \$10.00 & \$86.19 & \$1,093,085 \\ 
Health \& Fitness (101) & \$0.99 & \$69.99 & \$4.99 & \$13.28 & \$892,446 \\ 
Media \& Video (449) & \$0 & \$995 & \$4.99 & \$22.39 & \$6,568,106 \\ 
Music \& Audio (270) & \$0.99 & \$240 & \$4.99 & \$12.59 & \$2,723,183 \\ 
News \& Magazines (138) & \$4.99 & \$15 & \$8.99 & \$8.90 & \$62,659 \\ 
Photo \& Video (418) & \$0 & \$524 & \$4.99 & \$23.87 & \$5,781,482 \\ 
Photography (289) & \$0.99 & \$299 & \$4.99 & \$15.45 & \$1,710,315 \\ 
Productivity (301) & \$0.66 & \$524 & \$4.99 & \$21.57 & \$4,179,829 \\ 
Shopping (237) & \$5.99 & \$1,950 & \$46.87 & \$559.18 & \$370,268 \\ 
Social Media (26) & \$0 & \$10 & \$5.49 & \$5.99 & \$155,812 \\ 
Travel \& Local (171) & \$1.99 & \$49.99 & \$3.99 & \$15.99 & \$134,620 \\ 
Utility (1081) & \$0 & \$995 & \$4.99 & \$18.21 & \$6,456,385 \\ 
VPN (381) & \$2 & \$5,000 & \$5.99 & \$27.13 & \$2,159,982 \\ \hline
\textbf{Overall} & \$0 & \$5,000 & \$4.99 & \$24.41 & \textbf{\$40,644,171} \\ \hline
\end{tabular}}
\caption{Estimated cost of pirated software and subsequent financial damage}
\label{cost_analysis}
\end{table}

\subsection{Blackhat resources}
\begin{figure}[]
\centering
  \includegraphics[width=0.8\columnwidth]{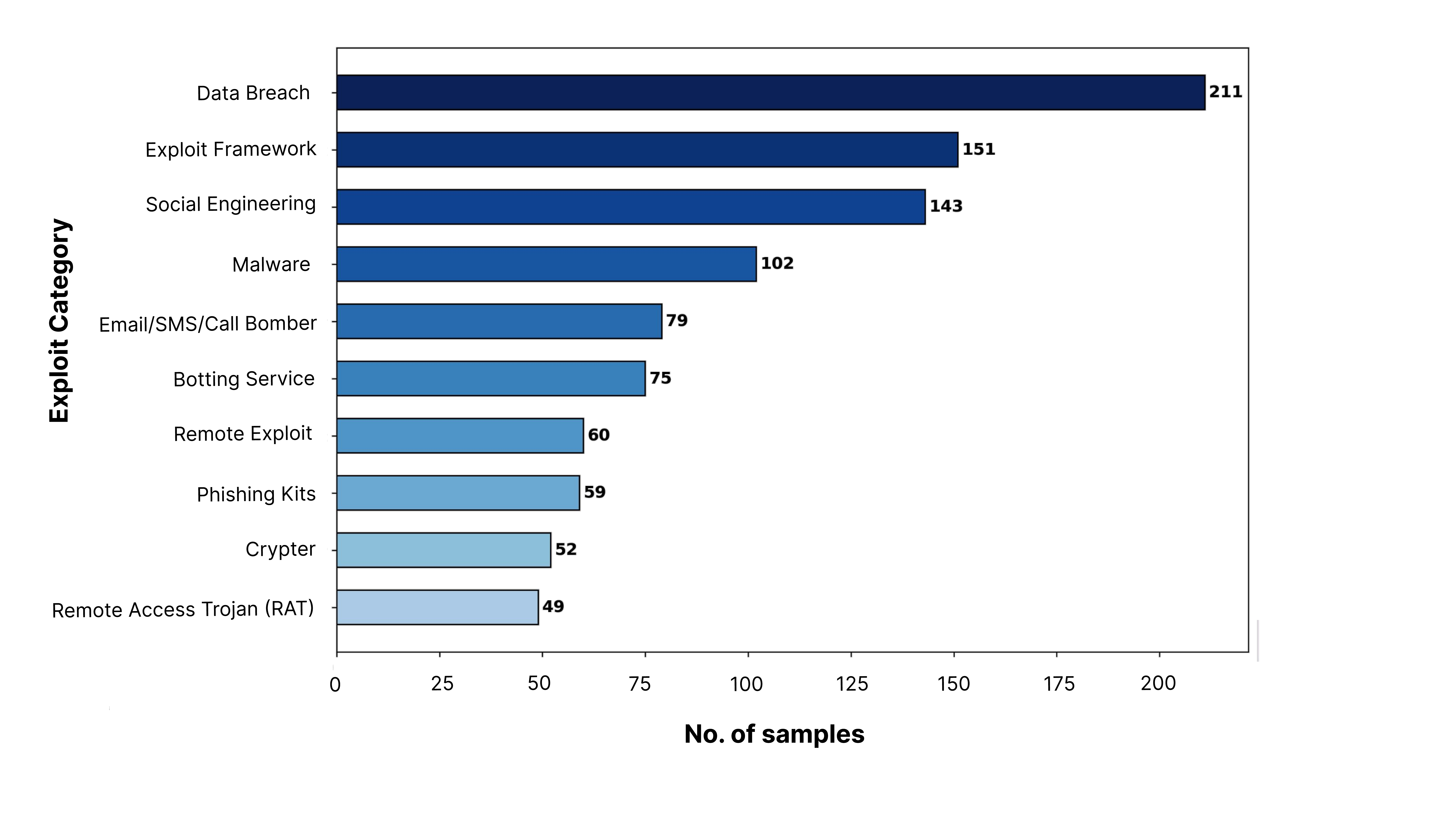}
\caption{Distribution of the ten most frequent exploit categories found in Blackhat resources CACs}
  \label{fig:top_10_exploits}
\end{figure}

We first manually reviewed 420 posts across 42 channels that shared resources (tools and training) for blackhat purposes. These materials ranged from basic hacking techniques to advanced exploitation methods.
Our classifier identified 2,541 posts overall across these channels. Notably, the post frequency in these channels was the lowest among all CACs, with a median posting interval of just over 27 hours. 
Due to the lower number of posts identified by the classifier, the two coders could manually examine the full sample of 2,541 posts. We found that 39\% of the posts (n=988), contained detailed, step-by-step instructions. Among these, 579 posts included YouTube links, while 135 posts embedded videos demonstrating how to set up and use specific hacking tools. These visual walkthroughs were presumably designed to make complex instructions more accessible to the audience. Additionally, 201 posts combined text instructions with images, often screenshots showing the tools in action, while 73 posts provided text-only instructions. These posts covered a range of blackhat hacking topics, including network penetration testing and the use of Remote Access Trojans (RATs).  Many posts promoted cybercriminal activities through practical demonstrations and real-world applications. For example, 15.9\% of the posts detailed successful phishing campaigns or provided instructions on executing Distributed Denial of Service (DDoS) attacks, with 59 posts offering the necessary toolkits. In 5.7\% of cases, these demonstrations were complemented by text or images, including success stories and testimonials from users claiming to have benefited from the shared knowledge. Other popular exploitation topics included instructions on setting up and managing botnets, spreading malware, using crypters (malware obfuscation), and cashing out cryptocurrency, offering aspiring cybercriminals a comprehensive toolkit. Similar to other CACs, nearly 12\% of the posts served as samples designed to encourage subscribers to purchase the complete tools required for an attack.
In contrast to GPT-4 used for categorization in the Pirated Software CACs (Section~\ref{pirated_software}, we used GPT-4 to confirm the categorization of the content within these channels that our coders had identified. Figure~\ref{fig:top_10_exploits} highlights the ten most frequently shared or discussed exploit categories. 
Beyond system exploitation tools, several tutorials focused on promoting social engineering tactics that extended beyond phishing. For example, 13.5\% of posts provided detailed instructions on impersonation and creating convincing fake profiles to deceive targets, often designed to bypass security measures and exploit human vulnerabilities~\cite{mink2022deepphish,tariq2022real}. Similarly, 7.2\% of the posts offered insights into email spamming techniques, promoting tools like SMTP settings and cracked email marketing software to facilitate unauthorized mass email campaigns.

We also identified 42 posts offering leaked user credentials for various email providers and online services. Similar to those found in Credential Compromise CACs, these also had similar forms of delivering the payload such as file attachments (25), external links (14) and through external bots/users (3) along with 16 posts that included "proofs" such as screenshots or videos demonstrating successful logins or transactions. 
% Eleven posts contained warnings about potential risks and provided guidance on avoiding detection by security systems, such as advising users not to use the same credentials across multiple IP addresses and to avoid accessing accounts from known devices or networks to prevent tracking.
We also reported the posts and corresponding channel names to Telegram and the targeted organizations, which included eight email providers (one exclusive to Credential Compromise channels), five domains, and seven online services (two exclusive) through their respective vulnerability disclosure programs.
These channels also shared technical insights and resources that, while valuable for legitimate cybersecurity purposes, could be repurposed for malicious intent. For example, 2.7\% of the posts discussed advanced vulnerabilities like Spectre and Meltdown, explaining their operation. However, these discussions also included tips on carrying out similar attacks, providing a resource that could be exploited by individuals with malicious intent. \shirin{it would be good to add an example.}

\subsection{Pirated media}
\begin{figure}[]
\centering
  \includegraphics[width=0.45\columnwidth]{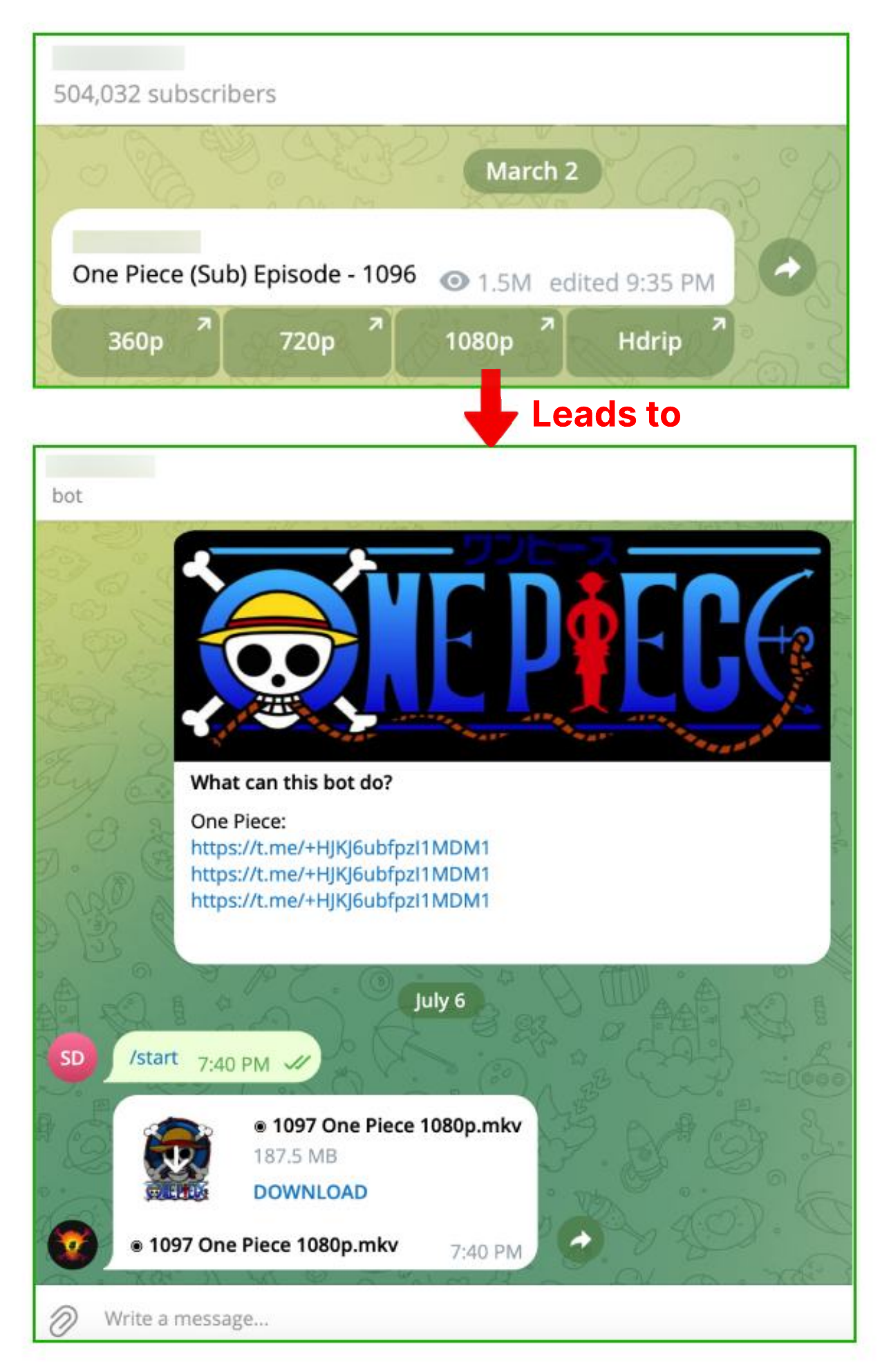}
\caption{Example of a Copyright media CAC which shares episodes of the One-Piece anime, asking users to engage with a bot to get the relevant media}
  \label{fig:pirated_media}
\end{figure}

We manually reviewed 360 posts from 36 channels that focused on distributing unauthorized copies of movies and TV shows, with our classifier finding a total of 2,039 such posts overall.
Much like the Pirated Software CACs, these channels primarily aim to provide free entertainment content to subscribers, typically including the title of a movie or TV show, a short synopsis or review, and technical details such as video resolution (720p, 1080p, or 4K).
Manually evaluating these posts, we found that, to deliver this content, channel administrators employ multiple distribution methods. In 34.6\% of posts, the media itself appears as a file attachment. Meanwhile, 42.3\% of the posts directed users to external websites where the content can be streamed or downloaded, and three channels consistently promote a single external site in every post. Additionally, 294 posts integrated “Content Access” or “Follow-to-Access” bots (discussed in Section \ref{bot_interactions}) that mediate file delivery, mirroring the bot strategies previously observed in Credential Compromise CACs. Figure \ref{fig:pirated_media} shows a sample from one channel that regularly uploads sequential episodes of the popular anime series "One Piece" \cite{onepieceIMDB}.
We also discovered that five channels concentrated on just one TV show each, uploading new episodes as soon as they were released. For ongoing series, fresh content appeared weekly, while for previously finished seasons, admins shared entire back catalogs at once. 

\subsection{Social media manipulation}
\label{social_media_manipulation}
\begin{figure*}[]
\centering
  \includegraphics[width=0.75\textwidth]{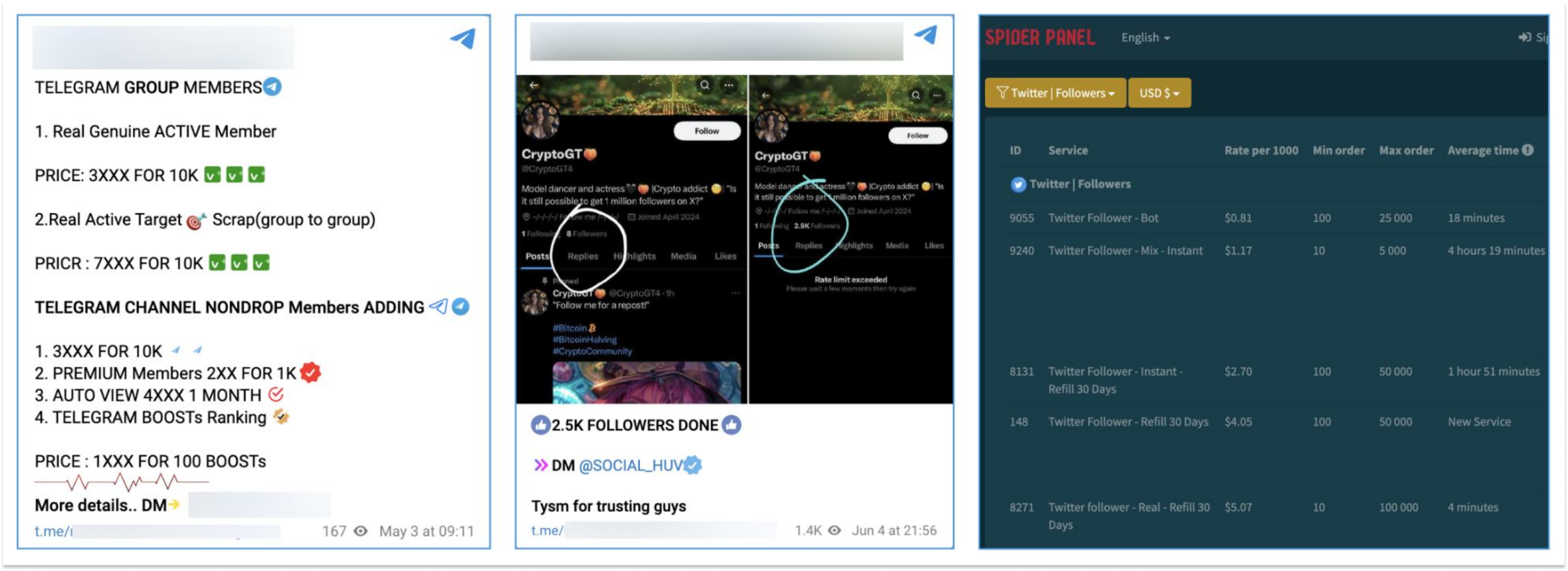}
\caption{Examples of artificial boosting: On left - Selling a fake account with significant engagement, On middle - Increase in followers post promotion, On right - A website where users can purchase artificial engagement}
  \label{boosting1}
\end{figure*}

We manually evaluated 680 posts from 68 Telegram channels advertising artificial engagement services—such as purchased likes, follows, or comments—for a variety of social media platforms. Extending this analysis via our automated classifier, we identified a total of 3,342 posts from these channels. 
To characterize and describe them in greater detail, we utilized the GPT-4 API, which revealed that these posts collectively targeted 57 platforms, spanning commonly used services like Twitter (now X), Facebook, Instagram, Telegram, WeChat, and Line, as well as more niche networks such as Peach, Vero, and Steemit.
Drawing on prior work by Nevado et al.~\cite{nevado2023analysis}, we found that these channels frequently referenced or promoted Social Media Marketing (SMM) panels - websites offering paid engagement (e.g., followers, likes, impressions) across multiple platforms. 
Overall, 4,051 unique websites appeared in these CACs designed to boost engagement metrics for 51 different platforms. 
\newline
\textbf{Methods of Building Trust and Advertising:} Beyond enabling the purchase of fake engagement, 39 channels shared images as "proof" of successful services, likely to build trust with potential customers. These images showcased tangible results, such as increased engagement or follower counts, demonstrating the effectiveness of the services provided. For instance, Figure~\ref{boosting1} highlights a channel displaying proof of a successful Instagram follower purchase, showing a visibly boosted follower count.
Additionally, 45 channels frequently hosted time-limited giveaway competitions, offering their services for free to winners. Notably, three channels accounted for a significant number of giveaway posts (n=480) featuring collaborations with social media influencers boasting substantial followings. These posts often guaranteed a specific number of new followers for giveaway winners. For example, one post promoted a giveaway by an influencer with 9.1 million followers, promising winners between 10,000 and 15,000 new followers. These giveaways likely served multiple purposes: attracting new users through risk-free trials, increasing the visibility and engagement of the hosting channels, and generating urgency and excitement around the services. Collaborations with popular influencers further amplified the perceived legitimacy and appeal of the services, benefiting both the influencers and the service providers. Such strategies align with findings in related research, such as the promotion of NFTs~\cite{roy2024unveiling}, where scammers exploit artificial promotion to lure unsuspecting victims with promises of high returns or  benefits. In Section~\ref{engagement_analysis}, we see how such "proofs" and giveaways contribute to enhanced engagement on these channels.
\newline
Taken together, these observations highlight a tightly coordinated ecosystem in which the CACs employ varied strategies—ranging from SMM panel promotions to bot-driven “funnel” workflows—to convince subscribers of the legitimacy and value of social media manipulation servers. Such behaviors not only inflate user metrics but also threaten the integrity of social media interactions at scale. 

\subsection{Telegram's reaction to channel disclosure}
\label{disclosure_and_removal}
We reported all 339 CACs to Telegram via their abuse email, including detailed evidence such as the channel name, URL, description, and summaries of the latest 10 malicious posts detected by our model. Evidence-based reporting, as opposed to generic submissions, has proven more effective in facilitating takedowns~\cite{oest2020phishtime}.
Despite these efforts, only 64 channels (19\%) were removed, primarily in categories such as pirated media (33 channels), pirated software (19 channels), and social media manipulation (10 channels). The higher removal rate for pirated software and media channels suggests that Telegram’s moderation policies may prioritize content involving unauthorized media distribution, potentially driven by pressure from media publishers~\cite{singhal2023sok}.
In contrast, only 2 out of 42 blackhat resource channels were removed, raising questions about whether these removals occurred independently of our reports and whether channels sharing exploit tools and hacking materials are subject to minimal scrutiny. Telegram’s moderation strategy appears to rely heavily on user reporting~\cite{wijermars2022telegram}. This reliance is reflected in subscriber-based visibility: pirated media channels average 137K subscribers, pirated software 84K, social media manipulation 11K (median), while blackhat resource channels only average only 3K subscribers. Channels with larger audiences are more visible, attracting greater scrutiny from both users and stakeholders~\cite{nilizadeh2016twitter,jha2024social}.

Conversely, Social media manipulation and Blackhat resource channels, with their smaller subscriber bases, receive fewer user reports and consequently less attention from Telegram’s moderation efforts. Moreover, as discussed in Section~\ref{engagement_analysis}, blackhat resource CACs often form tightly-knit communities, where subscribers are less inclined to report the channel. This insular structure likely contributes to the limited removal rates observed in this category. On the other hand, the median age of the 339 CACs was 237 days, suggesting that older, more established channels may be more resilient to removal compared to newer channels. This hesitancy in removing a significant majority of CACs highlights the challenges of moderating older, well-established channels. We explore this hypothesis in Section~\ref{darkgram}, where we compare the removal rates of 196 newer channels discovered by our framework with those of older channels.

\subsection{Overlap of content between Telegram CACs and cybercriminal forums}
\label{overal_of_content}
We also wanted to identify potential overlaps between the content shared on Telegram CACs and that found on traditional cybercriminal forums. To do this, we used Harvest~\cite{weichselbraun2020harvest}, a toolkit designed for parsing and collecting posts from online forums to collect a random sample of 107,852 forum posts shared between February 21st and May 29th from four prominent cybercriminal forums -  sinister.ly, cracked.to, nulled.to, and Blackhat World. Posts were collected at the root level of the forums, meaning the data was gathered randomly as new posts were shared without targeting specific sub-forums/sub-threads. Each post included the title and message content but excluded user statistics to ensure anonymity. Unlike prior studies that have relied on publicly available data dumps containing user and post information~\cite{buck2021didn}, we refrained from using such datasets due to the ethical ambiguity of handling user-level data. 
To examine potential overlaps, we randomly sampled 500 posts from each of the four forums. Two coders analyzed these posts to identify discussions that matched the types of content shared across the five Telegram CACs. 
Among the sampled posts, 581 did not share any similarities with content found on Telegram CACs. Instead, these posts included topics such as beginner hacking tutorials, general cybersecurity news, and forum-specific meta-discussions like rules and bans. For the remaining 1,419 posts, we identified 873 posts (43.6\%) discussing Credential Compromise, 498 posts (25\%) discussing Social Media Manipulation, but only 48 posts (2.4\%) discussing Blackhat Resources. Interestingly, we did not find any posts discussing Pirated Software/Media. This absence could be attributed to the fact that such content is typically hosted on external websites~\cite{cheng2013pirate,athey2013nature}, fraudulent app stores~\cite{malware_appstore}, or even legitimate platforms such as YouTube~\cite{chu2022behind}. Among the 673 Credential Compromise posts, we observed patterns similar to payloads shared in Telegram CACs. These posts often detailed the target and volume of leaked data and provided download links for the stolen credentials.  The 498 posts discussing social media manipulation also primarily advertised links to SMM (social media manipulation) panels, as seen in Telegram CACs. The small subset of 48 posts on Blackhat Resources included discussions on tools such as vulnerability scanners, credential stuffing scripts, and exploit kits. However, by further looking into the structure of the forums, we found many blackhat resources were confined to tiered access levels, such as VIP memberships, which required payments or forum reputation points, restricting our ability to study these posts in detail.
Credential compromise CACs on Telegram also shared 14,574 unique URLs to download credentials, and we checked if these external URLs appeared over the full dataset (i.e., 107,852 posts) and found that these forums posts contained 3,438 domains (around 23.5\%) that belonged to the same websites that were shared on Telegram. Similarly, Social Media Manipulation CACs on Telegram had shared 4,051 unique URLs linked to SMM panels, and we found 2,989 (around 74\%) of those domains appearing on the cybercriminal forums. 
This indicates a noteworthy overlap between the content shared between Credential Compromise and Social Media Manipulation content CACs and our cybercriminal forums dataset, and conversely. While this is an initial and limited evaluation, it would be intriguing to dedicate future work to the granular details of how content-sharing dynamics work between these two ecosystems.
\newline
\textbf{Annotation using Generative AI:} We used GPT-4 for annotating data from posts shared in "Credential Compromise" and "Social Media Manipulation" CACs, depending on its proven effectiveness in similar tasks as highlighted in prior work~\cite{gilardi2023chatgpt,tornberg2023chatgpt}. To ensure accuracy, we also manually evaluated a sample of the annotations in each instance. However, we acknowledge the reliance on a proprietary platform raises concerns about long-term replicability. In the event GPT-4 becomes unavailable or unsupported, the annotations it generated can be reproduced using open-source alternatives like LLaMA 3.1, which not only demonstrate comparable or superior performance to GPT-4~\cite{meta2025llama3} but also benefit from a more recent knowledge cut-off dates.

%%%%%%%%%%%%%%%% SECTION 6 %%%%%%%%%%%%%%%%%%%%%%%

\section{Subscriber harm analysis}
\label{harm_analysis}
In the previous section, we explored how CACs distribute malicious/illegal content that can be used by its subscribers to exploit third-party individuals and organizations. Here, we shift our focus on content that can instead \emph{pose a risk to the subscribers themselves}. With the volume of users subscribing to these CACs, we hypothesize that they can be exposed to content that can be malicious to them.  We examined 13,726 URLs and 3,205 executables that were shared in the CAC posts. Note that while phishing URLs were shared across all CACs, executables were chiefly (~97\%) found across Pirated software and Blackhat Resource CACs only. 
\newline
\textbf{Malicious URLs and Phishing Scams:}
Each URL shared in the CACs was scanned with VirusTotal~\cite{VirusTotalAPI:2020}, an online scanner that aggregates results from 80 anti-phishing engines. URLs flagged by two or more engines were marked as malicious, a threshold commonly used in prior research~\cite{salem2021maat}. This criteria led to 3,028 URLs being flagged as malicious. Recognizing VirusTotal's gaps, particularly with newer threats~\cite{vtpaper_blackbox}, we also scanned undetected URLs using PhishIntention~\cite{liu2022inferring}, a deep learning model for phishing detection based on website appearance and behavior. PhishIntention identified 829 additional phishing URLs, raising the total to 3,857, representing 28.1\% of all URLs shared in these channels. 
Pirated software and media channels contained the highest number of phishing URLs, with 1,507 and 1,110 flagged URLs, respectively, followed by Social-media manipulation channels having 638 and Credential compromise channels having 515 URLs. Prior literature~\sayak{cite}\cite{kumar2016malware,watters2021consumer} has found that users seeking unauthorized access to software or media are often less cautious about clicking on unverified links, making them prime targets for phishing attacks. Cybercriminals exploit this behavior by embedding phishing URLs within download pages, streaming sites, or advertisements. In contrast, Blackhat resource CACs contributed only 87 phishing URLs, as users in these spaces might typically possess greater technical expertise and demonstrate higher vigilance, as also found in similar groups in traditional cybercriminal forums~\sayak{cite}\cite{biswas2022text}. 
We reported all malicious URLs to five popular anti-phishing blocklists (Google Safe Browsing, APWG eCrime, PhishTank, OpenPhish, and Microsoft SmartScreen) and reported the posts containing these URLs to Telegram.
\newline
\textbf{Malicious files:}
\label{malware_harm_analysis}
To analyze executable files shared within CACs, particularly in Pirated Software channels, we used Hybrid Analysis~\cite{hybrid_analysis}, an online malware analysis platform developed by CrowdStrike and VirusTotal.  A file was considered malicious if detected by Hybrid Analysis and also had at least two antivirus scanners flagged it.  Using this approach, we found 1,210 files to be malicious, out of which only 491 (\~40\%) had been priorly scanned by Hybrid Analysis, suggesting several of the malicious files shared in the CACs had not been seen by the tool. Considering Hybrid Analysis is a popular tool that contributes threat intelligence to antivirus vendors, there is a possible detection gap for these files. Also, 310 of the malicious files were Android APKs and we cross-referenced them with those listed in the AndroZoo repository \cite{allix2016androzoo}, using package names to avoid discrepancies caused by modified file hashes. Interestingly, we found that 83 of the malicious APKs had corresponding entries available on the Google Play Store indicating Telegram’s role in distributing repackaged or potentially malicious apps.

\section{Channel Engagement}
\label{engagement_analysis}
In Section~\ref{characterization}, we established how the characteristics of various CACs with respect to content and how they are distributed. This section focuses on how users engage with this content by examining metrics such as follower growth, retention, and user reactions across the 339 CACs in our database throughout the study period. These channels collectively contained 23.8 million users, with a median of 8,000 followers per channel.  However, only channel administrators can access follower lists of a channel, which prevented us from determining whether individual users were engaging with multiple channels within the same CAC or across different CACs. 
\newline
% To compare the growth of these channels over time, we conducted a Wilcoxon's rank-sum test~\cite{woolson2007wilcoxon} between different CACs on a weekly basis. We analyzed the weekly growth in subscribers, comparing follower counts week over week, resulting in 20 observations over a 20-week period. Figure~\ref{fig:media_vs_weeks} in the Appendix shows a heatmap comparing the growth of Pirated media channels to other CACs over this period. It indicates that Pirated media channels have statistically significantly higher overall growth compared to other channels. For brevity, we only illustrate the comparison between media and other channels here. To summarize other comparisons: pirated software distribution channels exhibit significantly higher growth than the other three CACs; credential leaks outperform artificial boosting, and artificial boosting outpaces blackhat resources. \shirin{it's not clear why these results are important.}
\textbf{Impact of Proofs and Samples}
We examine how different strategies employed by channels, such as sharing proofs and samples, influence subscriber growth. As discussed in Section~\ref{characterization}, CACs such as Credential compromise and Social Media manipulation often share "proofs" to verify the legitimacy of their services or provide giveaways/samples for potential subscribers to try. We hypothesize that these strategies may drive subscriber growth, and thus compared the subscriber growth of channels that shared proofs with those that did not during the same period for both the CAC categories. While we did not find statistically significant growth immediately after sharing proofs, channels sharing proofs and samples had significantly higher overall growth than those that did not within a week ($p<0.01$ and $p<0.05$, respectively), suggesting that sharing samples or giveaways effectively attract new subscribers. Giveaways also often involve activities requiring participation, generating interest through word of mouth. Thus, while proof-sharing may not result in immediate subscriber increases, it also builds trust by demonstrating the service's credibility and attracting more subscribers over time.
\newline
% \subsubsection{Frequency of Posts}
% \shirin{Many papers show that reposting/number of posts affects account and content visibility. We should refer to them. Also, I am unsure how novel this analysis is and if it adds value to the paper.} 
% The \textit{frequency} of posts also impacts subscriber growth. To assess this, we measured the time intervals between consecutive posts from each channel and used this as a variable for a paired t-test. Statistically significant correlations between post frequency and subscriber growth were found in three CACs: copyright media, pirated software, and \shirin{remove?:} credential leaks. These channels tend to rely on frequent updates (e.g., new movies or TV shows for copyright media, fresh credentials for credential leaks, or frequent software updates for pirated software). In contrast, other CACs did not exhibit this trend. Engagement-boosting channels often share services or content that remain relevant over extended periods, diminishing the need for constant updates. Similarly, blackhat resource channels typically offer educational content with lasting value, as foundational techniques change slowly. This may explain why we did not observe a significant correlation between post frequency and subscriber growth for these CACs.
\textbf{Impact of Post Forwards}
Post forwarding is a critical driver of channel engagement on Telegram, enabling content to be shared across channels and significantly expanding its reach. To evaluate the impact of forwards on subscriber growth, we analyzed forward activity in CAC posts and correlated it with subscriber growth over a seven-day period.
Our findings reveal a strong positive association between the number of forwards and subscriber growth. Channels with posts that were forwarded more than 100 times in a week exhibited a median growth rate of 27.4\%, compared to 12.1\% for channels with fewer forwards ($p<0.01$). This trend was particularly pronounced in Pirated software and media channels. For instance, Pirated media channels with posts having over 100 forwards achieved a median growth rate of 39.2\%, compared to 17.3\% for channels with fewer forwards ($p<0.001$). Similarly, pirated software channels saw a median growth rate of 31.9\% for posts forwarded over 100 times, compared to 14.2\% for channels with lower forward activity ($p<0.001$).
The nature of content shared in these channels plays a pivotal role in driving forward activity. Pirated media channels often feature newly released materials that subscribers are eager to share, increasing the channel's visibility and attracting new followers. Similarly, Pirated software channels share links to high-demand software or updates, further amplifying their appeal and leading to substantial subscriber growth.
\newline
\textbf{Channel Migration:}
\label{channel_migration}
In Section~\ref{disclosure_and_removal}, we highlighted that 64 channels were deleted or removed by Telegram during our analysis period. Among these, 52 channels proactively posted alternate channel links to maintain their operations. Our analysis revealed a significant follower migration to these new channels, with a median of 43.8\% of followers transitioning within a week. This rapid migration likely reflects the reputation and trust these channels had previously established. When a well-known channel relocates, its audience is motivated to follow, ensuring continued access to the desired content.
This observation underscores the limitations of channel deletion as a standalone countermeasure. Cybercriminals can adapt quickly by creating and sharing alternate or backup channels, effectively retaining their follower base. Notably, channels dedicated to blackhat resources and credential leaks demonstrated the highest follower migration rates, at 65.3\% and 51.7\%, respectively. This suggests that subscribers to these channels are particularly dependent on the content provided, further emphasizing the resilience of cybercriminal activity in the face of takedown efforts.
\subsection{Emoji reactions}
\begin{figure}[]
\centering
  \includegraphics[width=0.85\columnwidth]{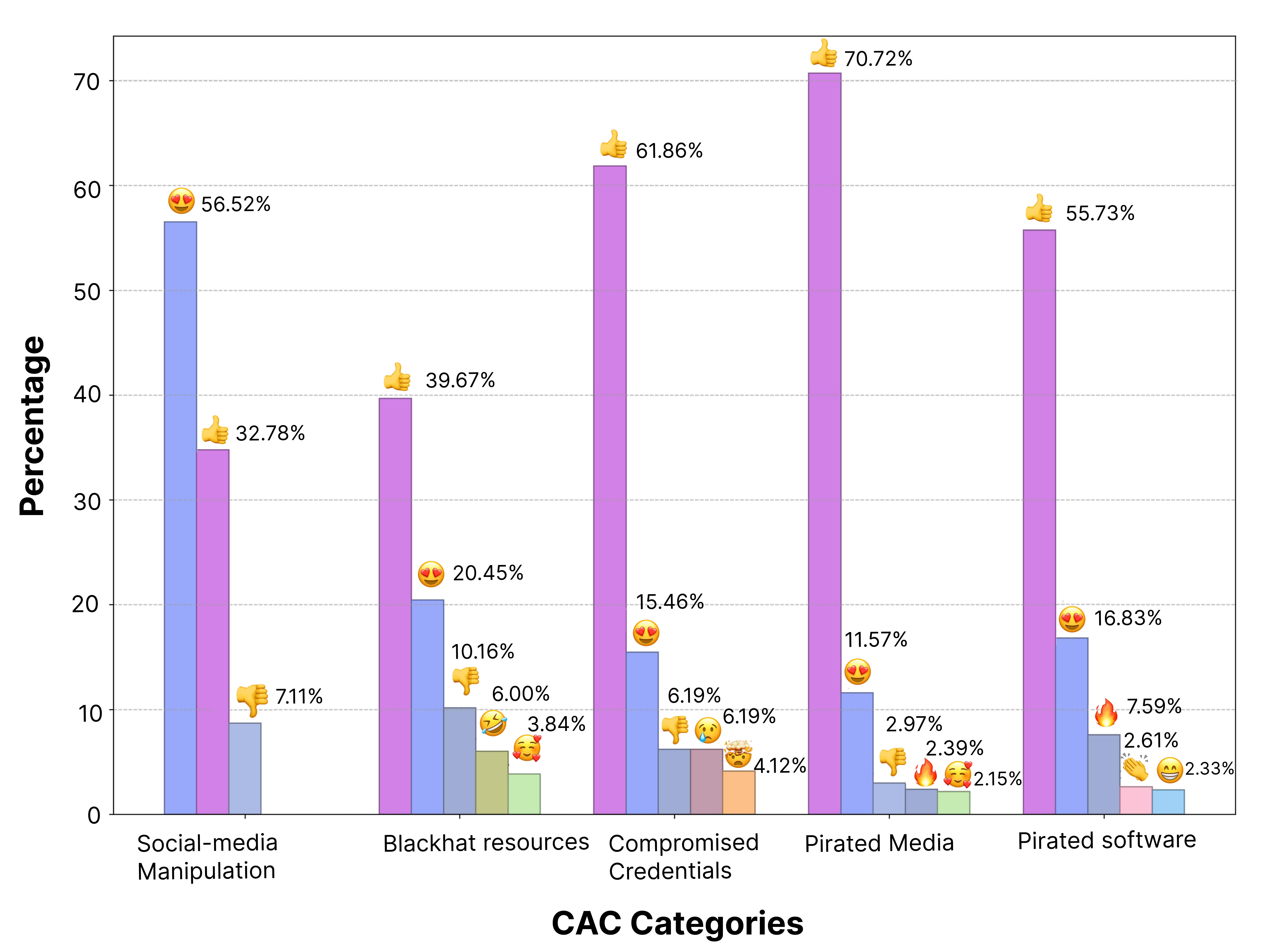}
\caption{Five most popular emojis per CAC}
  \label{fig:emoji_distribution}
\end{figure}

A majority of posts in the CAC channels (81.3\%) were either closed to replies or did not receive any, and the replies that did exist were typically very brief, with a median length of only 4 words and an average of 9.16 words. This brevity stands in stark contrast to the more detailed replies~\cite{gligoric2020adoption} commonly seen on other social media platforms like Twitter (now X) and Facebook, making interpretation more challenging. Therefore, we instead focus on interpreting the emoji reactions to the posts in this section.
Emojis provide a quick and expressive way for users to convey emotions and reactions without relying on lengthy text~\cite{riordan2017emojis}. In CACs, analyzing emoji reactions offers valuable insights into user engagement and sentiment toward shared content. Emojis can reflect a range of emotions, such as approval, excitement or disapproval. By analyzing the frequency and types of emojis used in response to CAC posts, we can better understand the effectiveness, popularity, and overall community sentiment to the shared content. 
\newline
\textbf{Analysis:} We found that only 22.8\% (n=32,480) of the posts had at least one reaction, with just 10.43\% (n=14,872) having more than one reaction. Figure~\ref{fig:emoji_distribution}
illustrates the five most frequently used emojis per CAC category. Notably, there is a lack of variation, with just ten emojis accounting for 74.8\% of all reactions out of the 68 unique emojis used in the dataset.
Overall, positive emojis such as "Like" and "Love" dominate reactions, suggesting that when users react to posts, their responses are generally positive. The "Like" emoji, which specifically indicates approval of the shared content, is the most frequent emoji across all CACs, except in Artificial Boosting, where it is the second most common. Similarly, the "Love" emoji, symbolizing fondness or happiness, ranks second in all four CACs. Other frequent positive emojis include the "Folded Hands" emoji (expressing gratitude), the "Fire" emoji (indicating excitement), and the "Exploding Head" emoji (signifying shock or amazement). These findings suggest that subscribers tend to approve of and engage positively with the content shared across all CACs. In Section~\ref{harm_analysis}, we explore whether subscriber reactions remain positive when harmful or malicious content is shared.
Conversely, two negative emojis, "Dislike" and "Crying Face," are somewhat prevalent across all CACs except Pirated Software. The "Dislike" emoji expresses disapproval, while the "Crying Face" emoji reflects sadness or distress. However, subscribers may simply ignore unsatisfactory content rather than react negatively.
\newline
\textbf{Reactions to Sales and Malicious Files:}
We also examined emoji reactions to posts promoting content for sale (e.g., through shared samples and links to external bots) and those distributing malicious files or phishing URLs. This analysis aimed to assess user satisfaction with the sales for the former and to gauge user awareness of the risks associated with the latter.
Similar to overall reactions to CAC posts, the majority of responses to both categories were overwhelmingly positive. For content-selling posts, 7 out of the top 10 emojis indicated interest or satisfaction, suggesting that users were engaged and generally pleased with the purchasing process.
On the other hand, posts sharing phishing links or malware also received predominantly approving reactions, with 9 out of the top 10 emojis reflecting consent or approval. This highlights a critical concern: many users may be unaware of the malicious nature of the content they are interacting with, inadvertently exposing themselves to significant risks. Such behavior underscores the urgent need for enhanced user education and awareness to mitigate these vulnerabilities.
\section{Interpreting cross-platform dynamics }
\label{sec:interpretation}
Our analysis reveals several behavioral and structural patterns that differ markedly from those commonly observed on traditional cybercriminal forums contextualized with the help of prior research into such forums.
\newline
\textbf{Accessibility to content:} Many cybercriminal forums are structured around multi-threaded, discussion-heavy ecosystems that rely on user-to-user interactions to foster trust, reputation, and mentoring~\cite{mcalaney2020knowledge,leukfeldt2017use}. Accessing relevant material often requires browsing subforums, navigating multiple contributors, and becoming familiar with specialized jargon. Some forums also impose barriers—such as invite-only registrations or tiered “VIP” access levels—to maintain the exclusivity and quality of shared content~\cite{leukfeldt2017use,pastrana2018crimebb}. 
We also specifically noticed this in Section~\ref{overal_of_content} where most blackhat resources on these forums where behind tiered barriers.
While these mechanisms help sustain high standards among seasoned members, newcomers can find access to content daunting. 
In stark contrast, Telegram’s CACs operate on a broadcast model, wherein administrators share content %(e.g., phishing kits, pirated software) 
via a simple chronological feed. 
Telegram’s CACs often present content in a format that requires minimal active engagement from subscribers. Additionally, each Telegram CAC typically focuses on a single type of content, streamlining the user experience by eliminating the need for extensive browsing or vetting. 
\newline
\textbf{Heightened Subscriber Risk:}
% Despite the easier accessibility of cybercriminal content on Telegram CACs, 
We argue that subscribers face potentially greater risks on Telegram CACs' channels than traditional cybercrime forums. Research on such forums reveals a surprising degree of self-regulation~\cite{dupont2022countering}, where experienced members act as informal gatekeepers by spot-checking questionable posts and leaving negative feedback to deter low-quality or harmful contributions~\cite{mcalaney2020knowledge} - mechanisms which help to mitigate risks for members.
In contrast, our analysis of Telegram CACs an environment which can be dangerous to the subscribers. Specifically, in Section~\ref{harm_analysis}, we find 28.1\% of URLs shared on these channels led to phishing sites, and 38\% of shared executable files were malicious. Unlike forums, where scams/harmful content typically unfold over days or weeks, allowing members to warn others~\cite{lu2010social}, administrators on Telegram CACs can inject such content into the feed almost instantly, and since the majority of the CACs disable replies to posts, wary subscribers have limited opportunity to raising alarms then they notice suspicious activity. It was also alarming that a significant portion of such harmful material is shared in channels focused on pirated software, media, and social media manipulation - content that is seeked by regular users~\cite{athey2013nature} instead of technically proficient hackers . This is further evidenced by our findings in Section~\ref{engagement_analysis} where users  often react positively to posts with malicious content, suggesting a lack of awareness of the risks in accessing them.
{\newline}
\textbf{Channel and Content Resilience:}
Cybercriminal forums are frequently taken down by law enforcement, the resulting fragmentation can dramatically reduce active membership{~\cite{pete2020social,samtani2020proactively}}. Many forums also depend on third-party FTP servers or specialized hosting providers for file storage, exacerbating the issue; if these hosts are shut down, the associated files vanish, even if the forum itself remains operational~\cite{mcalaney2020knowledge}. In contrast, Telegram CACs are more resilient. When threatened with removal, they simply create backup or alternate channels on the same platform and, as our findings indicate, can swiftly migrate a large portion of their subscribers, thereby preserving their community. Furthermore, the majority of cybercriminal content shared by Telegram CACs is hosted directly on the platform, eliminating reliance on external services and ensuring their content remains intact. CACs also often conceal shared content behind bots, likely making it more difficult for automated anti-scam crawlers to detect and remove it.
\newline
Telegram CACs also offer a more organized and seemingly “safer” means of distributing content that is otherwise scattered or easily taken down on other platforms. For instance, pirated software hosted on unreliable websites or malicious app stores often exposes users to malware and clickjacking \cite{la2021uncovering}. Similarly, pirated media is typically found on questionable streaming platforms or torrent sites, both of which are prone to adware, phishing, and takedowns\cite{lauinger2013clickonomics}. Even when such content is uploaded to legitimate streaming platforms like YouTube, it is often swiftly removed. By contrast, Telegram CACs centralize such content within a single channel, hosting it directly on the platform. From the end-user's perspective, this approach may provide a perceived sense of security, as downloading content from Telegram may feel safer compared to engaging with random, suspicious websites.
\newline
\textbf{Evolving strategies of monetization:}
Prior research has extensively examined monetization methods on cybercriminal forums\cite{zamani2019differences}, including middleman services, auctions, and escrow-like systems \cite{akyazi2021measuring, lummen2023telegram}. However, these markets are notoriously scam-prone \cite{bermudez2021shady}, featuring fake listings, impersonation of trusted sellers, or manipulation of escrow-like mechanisms—all of which erode trust among buyers and sellers. 
% Consequently, forums often rely on complex measures such as escrow services, middlemen, and anonymous payment portals to facilitate transactions in a landscape rife with scams. Even legitimate vendors resort to careful use of cryptocurrencies to evade law enforcement scrutiny.
In contrast, sales on Telegram CACs are exclusively controlled by channel administrators, who typically utilize bots (and, in some cases, affiliate users) to manage payments and deliver “premium” malicious content. This structure resembles a storefront environment rather than the distributed marketplace typical of forums\cite{leukfeldt2021cybercrimes}. Notably, posts advertising paid content on Telegram CACs routinely attract positive emoji reactions, an indication that buyers are generally interested or satisfied with their purchases, which is contrast to transactions in cybercriminal forums which can leave customers uncertain or dissatisfied~\cite{campobasso2023you}. Telegram’s built-in payment API further simplifies these transactions, removing the need for the complicated and anonymous payment portals that forum-based sellers utilize.
Another notable tactic on Telegram CACs is the use of “giveaways” and “proof” posts—practices seldom highlighted in existing forum-based research, to encourage user growth and reassure potential buyers. These curated “free samples” of stolen credentials or pirated software entice subscribers to pay for more. 
% In contrast, forums generally depend on a slower, more trust-oriented approach built on cumulative posts, user feedback, and verified transactions \cite{mcalaney2020knowledge, motoyama2011analysis}.
A significant challenge with Telegram CACs, however, is accountability. Whereas fraudulent transactions in forums can lead to public complaints and potentially result in moderators banning the perpetrator~\cite{hughes2024art}. Most Telegram channels also restrict user comments and hinder dissatisfied buyers from warning others. Consequently, victims’ only recourse is to report the channel to Telegram, which, as discussed in Section~\ref{disclosure_and_removal}, often appears reluctant to take action against well-established channels.
\newline
\textbf{Takeaway:} Prior literature has identified how compromised credentials{~\cite{bermudez2021shady}, SMM panels~\cite{nevado2023analysis} and blackhat resources{~\cite{samtani2020proactively} are regularly shared on cybercriminal forums. Our analysis in Section~\ref{overal_of_content} further reinforces these findings, revealing that forums not only exhibit similar payload characteristics but also rely on overlapping external websites to distribute their content. This highlights that Telegram CACs, at the very least, function as an alternative source for accessing cybercriminal content. Additionally, Telegram CACs share pirated content, which is largely absent from traditional cybercriminal forums. What sets Telegram apart is its ability to deliver this content in a streamlined manner that is readily accessible to a broad audience, and an apparent lack of strong moderation on the platform makes removal of such content less likely.
Our findings also reveal that this content not only reaches and is actively consumed by millions of users, not only necessitating the need for their swift takedown but also as a resource for security vendors for identifying malicious content. In the next section, we demonstrate how DarkGram, our BERT-based detection pipeline, can be extended to identify and disrupt newly emerging CACs in real-time.

% Thus, Telegram CACs have emerged as low-friction, broadcast-style hubs for cybercriminal activity, capable of delivering high volumes of malicious content directly to subscribers. This efficiency, coupled with accessible channel links, also poses risks for newcomers who gain one-click access to content that can harm others and themselves.
% Despite takedown attempts, administrators can quickly reconstitute a channel under a new name and migrate followers, underscoring the necessity for proactive detection strategies. Recent methodological advances in analyzing underground forums, such as Di Tizio et al.'s{~\cite{di2023graph}} has demonstrated that network structures can be leveraged for more effective detection by using centrality metrics to create stratified samples for training ML classifiers. In the next section, we demonstrate how DarkGram—our BERT-based detection pipeline—can be extended to identify and disrupt newly emerging CACs in real time, complementing the reactive takedown approach with a more preventive front line against Telegram-centric cybercrime.
\section{Detecting New CACs Using DarkGram}
\label{darkgram}
We extended the DarkGram framework to proactively identify new CACs shared on Telegram and Facebook from May 26 to August 11, 2024. Our goal was to detect these channels early, report them to Telegram for removal before they could gain significant subscribers, and notify affected stakeholders and security vendors regarding the malicious content shared within these channels. Similar to how we reported the original CACs in Section~\ref{characterization}, DarkGram also automatically reports to Telegram's abuse email with the channel description, CAC category, and screenshots of the post.
To identify new CACs on Telegram, we first searched for URLs starting with t.me shared across the 339 CACs, either as a post or through a bot. This approach was chosen because the channels in our ground truth already represent a curated, high-risk dataset, making it both resource-efficient and effective to focus on the interconnected networks commonly observed in cybercriminal activity~\cite{lau2014probabilistic}. As described in Section~\ref{engagement_analysis}, "Channel Migration," CACs often promote and encourage users to join similar channels. By leveraging this observed behavior, our approach can increase the speed and likelihood of detecting new emerging CACs. Nevertheless, DarkGram can just as easily be run on a random sample of channels with equal ease.
For each t.me link, DarkGram analyzes the 10 most recent posts to emulate the methodology we used to identify malicious CACs from Telemetrio while building the ground truth (Section~\ref{identifying-seed-channels}). 
Depending on the flagged content, DarkGram extracted URLs and executable files for further analysis using VirusTotal~\cite{VirusTotalAPI:2020} and PhishIntention. Channels with five or more flagged posts were classified as malicious and reported. Building the ground-truth in Section~\ref{identifying-seed-channels} also informed this threshold, as legitimate channels would occasionally contain malicious posts, likely shared accidentally by the administrators, a phenomenon not uncommon in other social media platform~\cite{balestrucci2020credulous}. By setting a higher threshold, we ensure that only channels demonstrating consistent malicious behavior are flagged, reducing the risk of false positives. Using this approach, DarkGram identified 127 malicious channels out of 245 examined links.  To demonstrate that DarkGram can also be used to find CACs on other social media platforms, we run it on Facebook, which is often used by cybercriminals to share malicious content~\cite{cao2014uncovering}.
To identify CACs on Facebook, we utilized the CrowdTangle API~\cite{fb_crowdtangle} (now replaced by the Meta Content Library). Through CrowdTangle’s global search functionality, DarkGram extracted t.me links embedded in posts from public groups. These links were evaluated using the same process as links found on Telegram CACs, looking through 4,191 links from 3,002 groups and flagged 69 (shared in 14 groups) as malicious, which were reported. 
\textbf{\newline{Misdetections}: }Manually evaluating the 245 channel links found from Telegram towards the end of our study (in August 2024), we did not find any false positives but noticed that an additional 19 CACs that were initially missed by DarkGram were sharing malicious content. Looking more closely, we found that DarkGram had missed all but one of these channels since initially they did not have any/so few posts which did not trigger the five malicious post threshold to be labeled as malicious. The same pattern was also seen for links obtained from Facebook, where while we did not find any false positives in the 69 channels identified, we also randomly sampled and verified 200 channels (from 4,191),finding 4 CACs which were also missed by DarkGram due to lack of posts.
\newline
\textbf{Characteristics:} Overall, the 196 channels shared on Telegram and Facebook combined were spread across all CAC categories we have studied: Credential Compromise (56), Pirated software (72), Artificial Boosting (17), Pirated Media (23), and Blackhat Resources (28). Figure~\ref{fig:new_channels} in Appendix illustrates the distribution of the new channels found by DarkGram over the three months.  
Manually evaluating a maximum of 10 randomly selected posts from each of the 196 newly discovered CACs, we found that the content shared across them closely mirrors the patterns observed in the 339 ground-truth CAC channels. These similarities span all CAC categories, including tactics employed by channel owners, such as using "proofs" and bots for content distribution. For brevity, we do not go into the detailed statistics of this content. However, this finding underscores the persistence of the malicious activities previously documented in this paper extending to the newer channels as well. 
Based on the creation dates of the channels, we found that the new channels had a median channel age of only 27 days compared to that of our original CAC channels at 237 days. This indicates that DarkGram was able to find significantly newer channels, as also justified by the newer channels having a median follower count of only 1,408.  
\newline
\textbf{Disclosure and Removal:} Reporting these channels to Telegram led to the removal of all 196 channels, with a median response time of 4 days. This is in contrast to when we reported the original channels in Section~\ref{disclosure_and_removal} where only 64 (of 339) channels were removed. With Telegram's content removal policy not transparent on how reported channels are moderated, we assume that since the original CACs were more established with higher subscriber counts and more traffic, Telegram might be more resistant towards removing them, as is also the practice on other social media platforms where scammers taking over popular accounts are more resilient than newer accounts~\sayak{cite}\cite{cobbe2021algorithmic}. This further brings into focus the importance of identifying and removing the CACs early. We also reported 20 of the 23 false negatives missed by DarkGram (19 from Telegram links and 4 from Facebook) at the end of the study. By that time, these channels had a median subscriber count of 8,125, significantly higher than when they were initially scanned by DarkGram. The three channels that were not reported, was since they had already been removed/takedown. Our disclosure resulted in the removal of 14 channels (as communicated by Telegram and verified by us), while 6 remained active.
It thus appears that Telegram's urgency in removing the channels becomes less severe the more popular the channel becomes, further emphasizing the need for early detection, and  DarkGram appears to have a positive impact in identifying and removing these channels. This is in line with recent reports of Telegram's struggles to stop cybercriminal activity from running rampant on their service~\cite{la2021uncovering, sreeram2024algorithmic}, while the reports focus on cybercriminal activity in private chats, through this work we see that malicious activity is just as functional on public easily accessible channels and often without any resistance from Telegram. 
We also reported 597 malicious URLs and 314 files, resulting in the removal of 343 URLs, 157 of which were confirmed by domain providers as a direct result of our reports. 
\section{Conclusion and Future Work}
We present the first comprehensive analysis of dedicated cybercriminal channels on Telegram, uncovering how the platform has become a hub for sharing malicious content across five distinct categories. Our findings highlight several key aspects: the unique characteristics of the content shared, the strategies employed by channel owners to attract users, build trust, monetize their operations, and ensure resilience against removal. These factors which differentiate these channels from traditional cybercriminal forums and also in several cases make them more potent. To address the threats posed by these channels, we developed DarkGram, a real-time framework capable of detecting and reporting cybercriminal channels and their malicious content. This proactive approach is critical given Telegram's apparent reluctance to remove older, more established channels. Our work also paves the way for future research solely dedicated to each CAC category, as the unique strategies and engagement methods merit deeper exploration to uncover more nuanced insights and inform targeted interventions. By open-sourcing our dataset and the DarkGram model, we provide a valuable foundation for continued investigation and collaboration in this area.

\section{Acknowledgements}
This paper is based upon work supported by the Comcast Innovation Fund and by the National Science Foundation grant number 2309318. Any opinions, findings, conclusions or recommendations expressed in this material are those of the author(s) and do not necessarily reflect the views of Comcast or the National Science Foundation.

\section*{Ethical Considerations}

As per the USENIX Security guidelines for ethical considerations, we have included this ethics report in accordance with the principles outlined in their call for papers and submission policies.
Our research on Telegram CACs was conducted with a thorough consideration of ethical implications, grounded in consequentialist and deontological ethics. Throughout the study, we prioritized transparency, privacy, and the potential impact on various stakeholders, including Telegram users, cybercrime victims, and the broader cybersecurity community.

\subsection*{Stakeholder Considerations}  
Key stakeholders include the users and administrators of cybercriminal channels, potential victims (such as organizations and individuals affected by criminal activities), law enforcement agencies, and Telegram as a platform. We took special care to avoid interacting with sensitive content and focused exclusively on metadata and filenames, ensuring no personally identifiable information (PII) was involved. When releasing datasets, we employed Microsoft Presidio~\cite{presidio}, an offline Python library that anonymizes sensitive text in data, to obfuscate all PII and excluded links to any compromised credentials. A dual manual review process by two coders, including the first author, further ensured the absence of PII. Data that could not be fully obfuscated was excluded from our analysis, and we promptly deleted such information from our experimental setup. Similarly, for malicious content and phishing URLs, we removed the data after completing the analysis. Furthermore, we respect the rights and autonomy of individuals who engaged in potentially illegal activities as well by obfuscating the identifiable channel information from all screenshots in the paper, and not including their names in the text as well.

\subsection*{Consequentialist Ethics}  
In alignment with the \textit{Beneficence} principle outlined in The Menlo Report~\cite{bailey2012menlo}, we sought to minimize harm while maximizing the benefits of our research. The primary long-term benefit lies in identifying harmful CACs and their content, potentially leading to their removal and reducing the spread of cybercriminal activity on Telegram. We recognize the risk of unintended consequences, such as criminals adapting to evade detection. To mitigate this, we limited the disclosure of specific vulnerabilities or methods in our DarkGram framework that could be exploited by malicious actors. 

\subsection*{Deontological Ethics} 
In line with the {Respect for Persons}principle, as previously mentoned in "Stakeholder Considerations", we ensured that no private or personally identifiable information (PII) was compromised during the course of our research. Also, our team avoided interacting with cybercriminals directly, and we did not engage in any actions that would violate the privacy or rights of the individuals involved. We did not download files that contained sensitive information (e.g., credentials) and focused our analysis strictly on the metadata and public information available from Telegram channels.  Since our research did not involve direct interaction with human subjects or sensitive personal information, our study was deemed not to require an Institutional Review Board (IRB) approval. 

\subsection*{Disclosure and Vulnerability Reporting}  
When vulnerabilities or malicious content were identified, we followed responsible disclosure practices by promptly reporting them within 12 hours to Telegram and affected organizations (e.g., email providers, and domain hosts). We adhered to responsible disclosure timelines, allowing organizations to address the threats (for as long as required since verifying some of our reports, such as compromised credentials or malware can be a time-consuming process) before publicizing any findings. Also our public reports contain only aggregated statistics without identifying specific organizations. In our open-sourced dataset, we exclude any unresolved disclosures that had not been confirmed by vendors, which does reduce the size of our dataset that was open-sourced. As a result of these efforts, 196 channels were taken down, and several phishing and malware URLs were mitigated.

\subsection*{Legal and Ethical Compliance}  
We ensured that our research complied with all applicable laws and respected platform terms of service, especially in our interactions with the Telegram API. We avoided conducting live experiments that could harm legitimate users or violate any platform’s policies.

\subsection*{Proactive Ethics Consideration}  
Ethical considerations were proactively integrated at every stage of our research process as mentioned in the proceeding paragraphs. From project design to data collection and analysis, we continuously evaluated potential harms and benefits, allowing us to anticipate ethical challenges and implement solutions accordingly.

\subsection*{Open Science Policy}
To adhere with Open Science Policy, we open-source our dataset and also make our detection framework - DarkGram available at \url{https://zenodo.org/records/14736879}.
% \subsection*{Conclusion}  
% Our research contributes significantly to the fight against cybercrime on Telegram, and we are committed to upholding ethical standards that minimize harm, protect privacy, and respect individual rights qne continue to responsibly disclosure our findings to support ongoing efforts to secure online platforms.

% trigger a \newpage just before the given reference
% number - used to balance the columns on the last page
% adjust value as needed - may need to be readjusted if
% the document is modified later
%\IEEEtriggeratref{8}
% The "triggered" command can be changed if desired:
%\IEEEtriggercmd{\enlargethispage{-5in}}

% references section

% can use a bibliography generated by BibTeX as a .bbl file
% BibTeX documentation can be easily obtained at:
% http://www.ctan.org/tex-archive/biblio/bibtex/contrib/doc/
% The IEEEtran BibTeX style support page is at:
% http://www.michaelshell.org/tex/ieeetran/bibtex/
\bibliographystyle{plain}
\bibliography{refs}
\section{Appendix}
\subsection{GPT-4 prompts for categorization/labelling CACs}
\label{prompt_categorization}
This section highlights the prompts that we used for the categorization and labeling of posts in Pirated Software (Section ~\ref{pirated_software}) and Social-media manipulation (Section ~\ref{social_media_manipulation}) CACs using GPT-4 (OpenAI API version 2024-02-01 on Azure).
\begin{tcolorbox}[colback=blue!5!white, colframe=blue!75!black, title= Categorization in Pirated Software CACs]
\footnotesize
This message: \textbf{\{message\}} was shared in a Telegram channel known for distributing software. [Optional: The message also included an attachment with the filename \textbf{\{filename\}}.]

Based on the content of the message [optional: and the provided filename], please assign the referenced software to one of the following 15 categories: \textbf{\{categories\}}. If it does not fit into any of these predefined categories, feel free to propose a new one.

After categorizing the software, kindly provide a brief description of its functionality or purpose. Please \textbf{ONLY} output your response as:
\texttt{\{Software\_name, Category, Description of Software\}}
and no additional information.
\end{tcolorbox}

\begin{tcolorbox}[colback=blue!5!white, colframe=blue!75!black, title=Identifying app price in Pirated Software CACs]
\footnotesize
This message: \textbf{\{message\}} was shared in a Telegram channel known for distributing software. [Optional: The message also included an attachment with the filename \textbf{\{filename\}}.]

Based on the content of the message [and the optional filename], find the price of the software. If the software involves in-app purchases or subscriptions, determine the subscription cost and use it as the price AND also categorize it as Freemium or Premium. 

Please \textbf{ALWAYS} provide a US dollar value and \textbf{ONLY} output your response as \texttt{<price><category>} and no additional information.
\end{tcolorbox}

\begin{tcolorbox}[colback=blue!5!white, colframe=blue!75!black, title=Identifying the platforms targetted by posts in Social Media Manipulation CACs]
\footnotesize
This message: \textbf{\{message\}} was shared in a Telegram channel known for prompting services that artificially inflate social media metrics (such as likes, followers, shares, etc.).

Based on the content of the message, identify which particular social media platform it is targeting.

Please \textbf{ONLY} output your response in the format: \texttt{\{platform\}} and no additional information.
\end{tcolorbox}

\subsection{Bots shared by CACs} 
\label{bot_interactions}
We identified 71 distinct bots that were “tagged” in 64 credential compromise CACs as a medium to provide the compromised credentials to the user. Of these, 59 mentioned only a single bot, while the remaining five referenced two or more. We observed no cross-channel reuse of bots; each channel relied on a distinct bot (or set of bots), suggesting that channel administrators tend to adopt or develop their own dedicated automation strategies.
Upon interacting with these bots, we discovered three main categories based on their functionality and how they facilitate access to the compromised credentials:
\newline
\textbf{Payment gateway bots:} In 31 cases, the bot’s primary role was to process transactions using the Telegram Payment API~\cite{telegram_payment_api}. This feature allows channel administrators to integrate a payment interface directly into the bot, enabling subscribers to pay for premium content without leaving Telegram. Once a subscriber completes payment, it is assumed that the bot delivers the promised content. These bots act as a storefront for the CACs to facilitate transactions, eliminating the need for external escrow or manual negotiation found in traditional cybercriminal forums~\cite{uren2024telegram}.
\newline
\textbf{Content Access Bots:} 19 bots automatically granted subscribers access to files or download links without requiring any additional payment steps. For instance, a channel might post a brief description of a leaked database or a pirated software crack, followed by a link to a bot, upon accessing which the advertised file is provided. By routing downloads through a separate bot, channel administrators effectively keep the malicious payload off the primary CAC feed, where Telegram’s moderation or automated filters might detect it—instead, the bot acts as an indirect distribution hub.
\newline
\textbf{Follow-to-Access Bots:} 34 bots enforced a “funnel” mechanism, compelling subscribers to join or follow additional (usually 3 or more) channels before granting access to the payload. These channels were \emph{always} other credential compromise CACs. By compelling users to follow multiple other channels before unlocking the payload, these bots effectively allow similar channels in the ecosystem to gain followers, indicating that rather than operating as isolated entities, these CACs form a collective ecosystem where channel administrators collaborate—intentionally or otherwise to broaden their audience and strengthen their community presence. A future dedicated study to this dynamic is required.
\newline
\textbf{Characterizing bots shared by other CACs:}
In Pirated Software CACs, we analyzed 309 posts across 44 channels, identifying 51 unique bots facilitating access to pirated software. 22 bots were classified as \emph{Content Access Bots} providing direct downloads, while 29 bots functioned as Follow-to-Access Bots, requiring users to join other related channels before granting access. 
Notably, there were no Payment Gateway Bots in this category, as these channels focus on freely distributing pirated software.
On the other hand, the Social Media Manipulation CACs had 412 posts across 56 channels referencing 48 distinct bots. Of these, 23 were Payment Gateway Bots, facilitating cryptocurrency transactions for purchasing fake social media engagement, such as likes, followers, or comments. Another 18 bots were Content Access Bots, offering tools or services for social media manipulation directly to users. 
The remaining 7 bots employed a Follow-to-Access mechanism, compelling users to join related channels before granting access to tools or services. 
This combination of strategies reflects the dual focus of these channels on generating revenue and expanding their influence.
\newline
In the Pirated Media CACs, our analysis covered 294 posts across 39 channels, uncovering 25 unique bots, out of which 15 were Follow-to-Access Bots, while the remaining 10 bots were Content Access Bots, directly offering downloads of the advertised content. Similar to the Pirated Software CACs, no Payment Gateway Bots were present, as these channels prioritize audience growth over monetization. Finally, in the Blackhat Resources CACs we found 157 posts across 8 channels, we identified 10 bots, with 6 allow cryptocurrency transactions. While the remaining 4 bots included 3 Content Access Bots and 1 Follow-to-Access bot.

\begin{figure}[H]
\centering
  \includegraphics[width=\columnwidth]{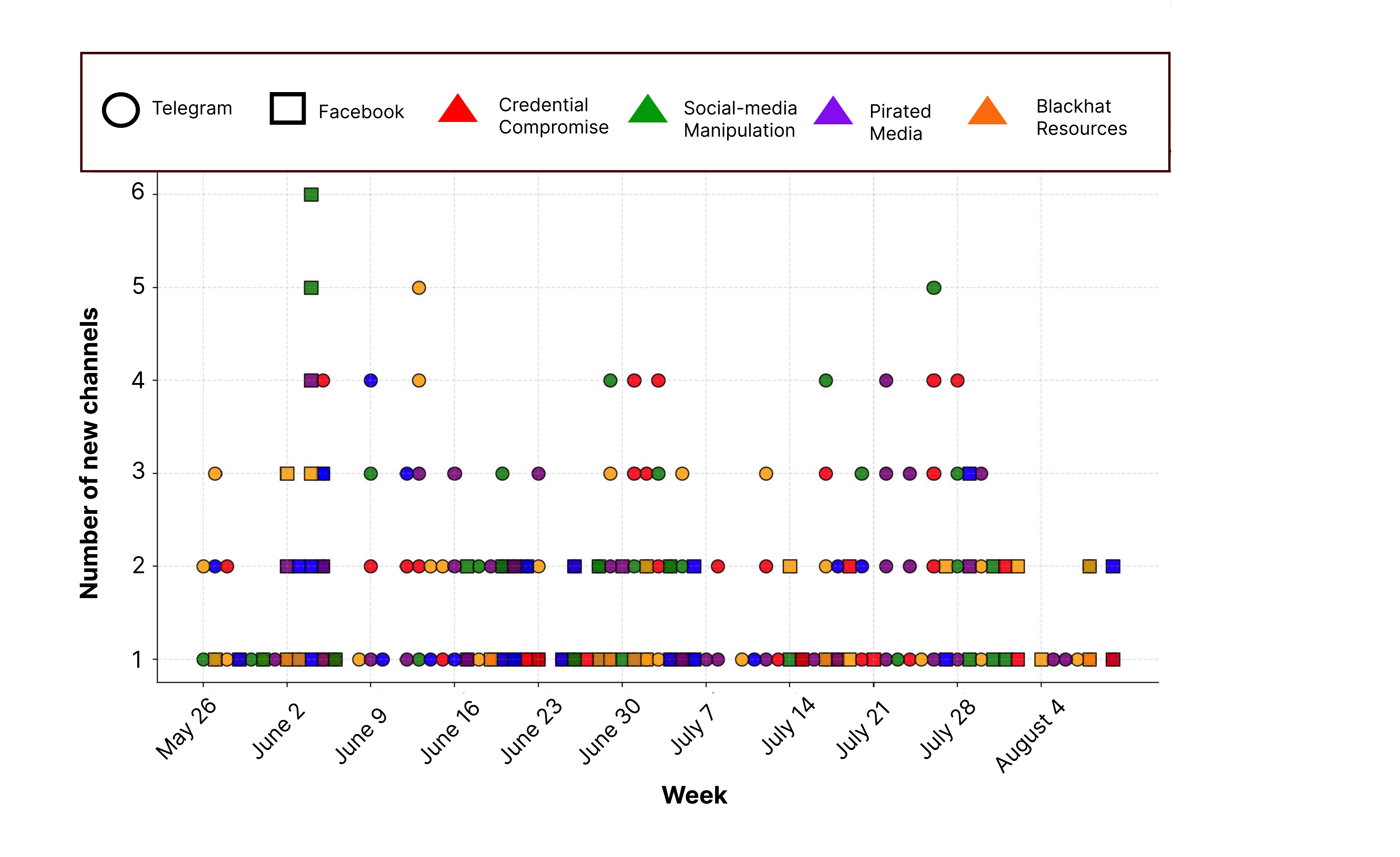}
\caption{Distribution of new CACs identified by DarkGram on Telegram and Facebook on a weekly basis}
  \label{fig:new_channels}
\end{figure}

% that's all folks
\end{document}